\documentclass[aps,prb,twocolumn,superscriptaddress,showpacs,floatfix,amsmath,amssymb]{revtex4}
\usepackage{graphicx}
\usepackage{amssymb}
\usepackage{color}
\usepackage{amsfonts}

\def\t6as{$\mathrm{(TMTSF)_{2}AsF_{6}}$\ }

\def\tmc{$\mathrm{(TMTSF)_{2}ClO_{4}}$\ }

\def\tms{$\mathrm{(TMTSF)_{2}AsF_{6(1-x)}SbF_{6x}}$\ }

\def\tmttfsbf6{$\mathrm{(TMTTF)_{2}SbF_{6}}$\ }

\def\tmttfasf6{$\mathrm{(TMTTF)_{2}AsF_{6}}$\ }
\def\tmttfbf4{$\mathrm{(TMTTF)_{2}BF_{4}}$\ }

\def\tmtsfreo4{$\mathrm{(TMTSF)_{2}ReO_{4}}$\ }
\def\tmno3{$\mathrm{(TMTSF)_{2}NO_{3}}$\ }
\def\tm2x{$\mathrm{(TM)_{2}X}$\ }
\def\tm2xns{$\mathrm{(TM)_{2}X}$}

\def\hc2{$H_{\mathrm{c2}}$\ }

\def\tmp6{$\mathrm{(TMTSF)_{2}PF_{6}}$\ }

\def\tms2x{$\mathrm{(TMTSF)_{2}}X$}
\def\tm2x{$\mathrm{(TM)_{2}}X$\ }

\def\4fb{$\mathrm{BF_{4}}$}

\def\reo4{$\mathrm{ReO_{4}}$}
\def\bedtttfreo4{$\mathrm{(BEDT-TTF)_{2}ReO_{4}}$\ }
\def\et2i3{$\mathrm{(ET)_{2}I_{3}}$\,}
\def\et2x{$\mathrm{(ET)_{2}X}$\,}
\def\ket2x{$\mathrm{\kappa-(ET)_{2}X}$\,}

\def\ket2x{$\mathrm{\kappa-(ET)_{2}X}$\,}

\def\betsfecl4{$\mathrm{(BETS)_{2}FeCl_{4}}$\,}

\def\et{$\mathrm{ET}$\,}

\begin{document}

\title{Superconducting and spin-density wave phases probed by scanning tunneling spectroscopy in the organic conductor \tmc}

\author{Mohammadmehdi Torkzadeh}
\affiliation{Sorbonne Universit\'e, CNRS, Institut des Nanosciences de Paris, UMR7588, F-75252 Paris, France}
\author{Pascale Senzier}
\affiliation{Laboratoire de Physique des Solides (UMR 8502) - Universit{\'{e}} Paris-Saclay, F-91405 Orsay, France}
\author{Claude Bourbonnais*}
\affiliation{Regroupement Qu\'eb\'ecois sur les Mat\'eriaux de Pointe and Institut Quantique,
D\'epartement de physique, Universit\'e de Sherbrooke,
Sherbrooke, Qu\'ebec, Canada, J1K-2R1}
\author{Abdelouahab Sedeki}
\affiliation{Universit\'e Moulay Tahar de Saida, B.P. 138 cité ENNASR 20000, Saida, Algeria}
\author{C\'ecile M\'eziere}
\affiliation{Univ Angers, CNRS, MOLTECH-Anjou, SFR MATRIX, F-49000 Angers, France}
\author{Marie~Herv\'e}
\author{Francois~Debontridder}
\author{Pascal~David}
\author{Tristan Cren}
\affiliation{Sorbonne Universit\'e, CNRS, Institut des Nanosciences de Paris, UMR7588, F-75252 Paris, France}
\author{Claire Marrache-Kikuchi}
\affiliation{Universit\'e Paris-Saclay, CNRS/IN2P3, IJCLab, 91405 Orsay, France}
\author{Denis~Jerome}
\affiliation{Laboratoire de Physique des Solides (UMR 8502) - Universit{\'{e}} Paris-Saclay, F-91405 Orsay, France}
\author{Christophe Brun*}
\affiliation{Sorbonne Universit\'e, CNRS, Institut des Nanosciences de Paris, UMR7588, F-75252 Paris, France}

\date{\today}

\begin{abstract}
By scanning tunneling microscopy (STM) we have probed the local quasi-particle density of states (DOS) of the Bechgaard salt organic superconductor \tmc in slowly cooled single crystals cleaved under ultrahigh vacuum conditions. In well STM imaged  crystallographic surface planes, the local DOS has been probed for different surface areas at temperatures above and below the critical temperature of superconducting or insulating spin-density wave states. While a rather homogeneous superconducting state is expected in the bulk from previous studies, depending on the degree of disorder introduced by cleavage in the anion lattice, an inhomogeneous granular state is predominantly observed at the surface. A pronounced linear V-shape profile of the local DOS is observed from intermediate  to the lowest energy scale in the  less disordered superconducting surface areas. This supports the existence of an unconventional d-wave like order parameter with nodes at low energy, which is preceded by more energetic fluctuations attributed to quantum criticality of the material. At higher energy disorder combined to correlations deplete further the DOS. By contrast a non-linear U-shape characterizes the local low energy DOS profile for the more disordered and insulating surface areas of the spin-density wave state. The  experimental results are compared quantitatively with those predicted by the renormalization group theory  of the quasi-one dimensional electron gas model and its description of the superconducting and spin-density wave states that are interlinked by quantum criticality in the Bechgaard salts.

\end{abstract}

\maketitle

\section{Introduction}
Among the various families of  unconventional superconductors discovered over the last four decades or so, organic materials occupy a  noticeable place.  This is  well exemplified by the 
first  organic superconductors to be synthesized, the Bechgaard salts [(TMTSF)$_2 X$] series.  These strongly anisotropic, quasi-one dimensional (quasi-1D), conductors are   distinctive by the emergence 
of  superconductivity   on the edge of an insulating antiferromagnetic spin density-wave (SDW)  state    
when  pressure is tuned  either hydrostatically or chemically 
from anion $X$ substitution \cite{Jerome24,Brown15,Jerome16}. This  sequence is  now known as a classical example of pressure-tuned antiferromagnetic quantum critical point (QCP)  made unstable by  the development of a superconducting dome. 

A major effort in the quest of understanding these materials has resided in explaining how both quantum criticality and superconductivity are reciprocally  conditioned.   To this end most sophisticated experiments  have been conducted on (TMTSF)$_2$ClO$_4$,  the only member of the (TMTSF)$_2 X$ series displaying at ambient pressure a superconducting state below $T_c=1.2$K,  which is  located  in the immediate vicinity of  the QCP.  Anomalous-featured order parameter  for  superconductivity is illustrated in various situations. These are for instance given  by the detrimental effect of non magnetic impurities on $T_c$ 
\cite{Joo05,Choi82}; the  power law temperature dependence of  the NMR relaxation rate \cite{Takigawa87,Shinagawa07} and of specific heat below $T_c$ \cite{Yonezawa12};  the  Fermi surface gap profiling  by angle-resolved field dependent specific heat and its $\sqrt{H}$ field profile   in the vortex state \cite{Yonezawa12}. Combined  with the suppression of the NMR Knight shift   in the superconducting state at low field \cite{Shinagawa07}, these findings support the existence of a singlet  order parameter for superconductivity with a nodal  structure on the Fermi surface, probably $d-$wave. Other experiments like  former specific heat measurements, thermal conductivity and $\mu$-sr, however,  do not accord with this view and  rather support  the existence of a nodeless  gap along  the  Fermi surface \cite{Garoche82,Belin97,Pratt13}.   

The hallmarks  of  quantum criticality fanning out through the  metallic state are found in different conditions.    This is the case of  the linear temperature dependence of electrical resistivity \cite{DoironLeyraud09,DoironLeyraud10}. Also  known as Planckian dissipation \cite{Legros19,Bruin13}, it is present over  almost two decades in temperature around $T_c$ in the Bechgaard salts. Another imprint comes from  NMR spin-lattice relaxation rate. Different set of experiments  reveal the existence of pronounced antiferromagnetic spin fluctuations, extending  from the lowest temperature to  far beyond $T_c$ in the metallic state and whose amplitude   is   tuned  by the pressure distance from the QCP \cite{Creuzet87b,Bourbon84,Brown15,Shinagawa07,Kimura11,Bourbon09}.

None of these observations address directly the spectroscopic properties of quasi-particles in the quantum critical metallic domain and how these modify when entering in the superconducting phase. It is the  main motivation of the present work  to address these important issues from the angle of tunneling spectroscopy that can probe  the quasi-particle density-of-states in the metallic, superconducting  and insulating phases  \cite{Wolf2011}.

\begin{figure*}[tbh]\begin{center}\includegraphics[width=\textwidth]{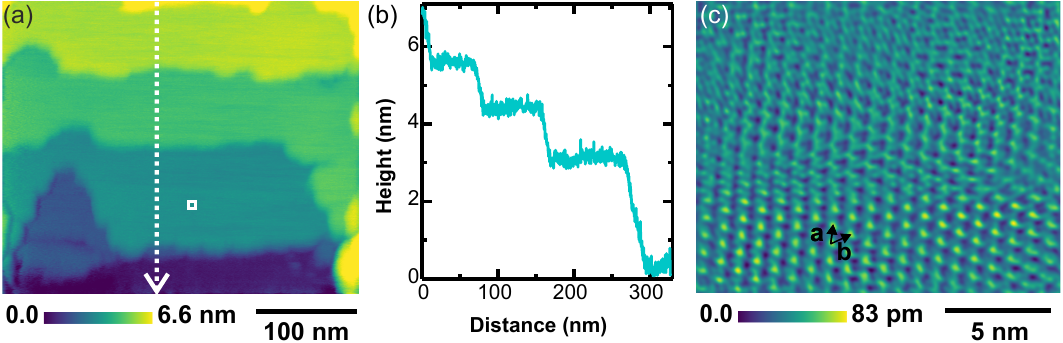}\caption{(color online) Measurements performed at $T=300$~mK. a) Raw color-coded constant current $Z(x,y)$ STM topography of a \tmc single-crystal cleaved under ultrahigh vacuum. The scanned area is $400\times330~\mathrm{nm}^2$ measured with $I=$50~pA and $V_{\mathrm{bias}}=-0.1$~V. Several crystallographic (\textbf{a},\textbf{b}) planes are seen. The color bar of $Z(x,y)$ expressed in nm is shown on the right. b) $Z(Y)$ profile of the STM tip acquired along the vertical dashed line shown in image a). Both displacements $Z$ and $Y$ are in nm. The height variations are consistent with the presence of elementary crystallographic steps along \textbf{c}. c) STM topography of a smaller area (of about $15\times15~\mathrm{nm}^2$), where molecular resolution could be obtained emphasizing the quasi-1D chains lying along \textbf{a}. The \textbf{b} direction is also indicated. The color bar of $Z(x,y)$ expressed in pm shows a corrugation less than 0.1~nm. This image was Fourier filtered to reduce the signal to noise ratio and highlight the molecular resolution. \label{fig1_STM_topos}}\end{center}\end{figure*}

Several attempts  to this end have been  performed in the past using different tunneling setups. Early experiments used a $n$-doped GaSb Schottky barrier evaporated onto a \tmp6 single crystal down to 50~mK under a pressure of 11 kbar \cite{More81}. It provided evidence of a well developed gap $2\Delta\approx 3.6$~meV at the Fermi level, interpreted as a superconducting gap, persisting above $T_c$ like a pseudo gap structure. Subsequent \tmc studies at ambient pressure revealed similar results for superconductivity and evidenced also the role of the cooling speed on ClO$_4$ anion disorder leading  to a shift on the pressure axis and to the onset of an insulating  SDW state.  An estimation of the SDW gap of $2\Delta_{\rm SDW}\approx6$~meV was given \cite{Fournel83}. Further planar tunneling SIS' experiment using \tmc-amorphous Si-Pb junctions, revealed a much smaller superconducting energy gap in the ordered state of $2\Delta \approx 0.8$~meV and an SDW gap of $2\Delta_{\rm SDW}\approx 3.0$~meV for the anion disordered state \cite{Bando85}. In \tmp6 at ambient pressure, a scanning tunneling spectroscopy study of the as-grown \textit{a-b} surface reported a SDW gap $2\Delta_{\rm SDW}\approx 4.6$~meV below the ordering temperature $T_{\rm SDW}\simeq 12$~K, and preceded over a broad range of  temperature by a pseudogap or dip in the density of states.  \cite{ICHIMURA2001}.

Given the diversity of these  results it turned out to be important to have a reliable spectroscopic tunneling method to perform a direct determination of the local quasi-particle density of states (LDOS) in the superconducting, metallic and SDW phases. The method we have used is very low temperature scanning tunneling microscopy/spectroscopy (STM/STS) after having cleaved the \tmc sample under ultrahigh vacuum. This technique has been widely used on inorganic conventional and unconventional superconducting materials \cite{Fischer2007}, including layered organic superconductors \cite{Guterding16,Diehl15}; it  also has proven to achieve molecular resolution when applied to Q1D organic charge transfer compounds \cite{Wang03}. 

In this work, we could for the first time succeed to image the elementary (\textbf{a},\textbf{b}) planes of (TMTSF)$_2$ClO$_4$ resulting from the in-situ cleavage and perform STS on these planes.  The $dI/dV$ tunneling conductance measurements performed at $T=300$~mK revealed various types of regions that are consistent with either superconducting or SDW phases depending on disorder conditions. Due to the cleavage process inducing disorder in the ClO$_4$ lattice sites, superconducting regions are rarely found in the surface plane. Instead, an inhomogeneous granular state is predominantly observed at the surface. In rare surface superconducting regions, the quasi-particle excitation spectrum in the meV energy range reveals a strongly V-shaped DOS presenting a finite residual DOS at the Fermi level. The shape of the local excitation spectrum measured by STS is consistent with  an order parameter having a nodal d-wave structure rather than  nodeless $s$-wave. In contrast SDW puddles reveal U-shaped excitation spectra consistent with a nodeless energy gap having larger values than superconducting regions. In the  energy range above the superconducting  gap and  above $T_c$, the V-shaped depletion of the quasi-particle  DOS persists  up to intermediate energy where disorder and correlations govern the depletion of the DOS. In the few meV range, we show that our results are rather congruent  with the predictions of the renormalization group approach to  the  quasi-1D electron gas model for the SDW to $d$-wave SC sequence of ground states exhibited across a QCP  \cite{Sedeki12}.

\section{Experimental set-up}
The STM/STS experiments were carried out in an ultra-high vacuum set-up having a base pressure in the low $10^{-11}$~mbar range. \tmc single crystals of about several mm lengths were selected and glued on stainless steel sample holders using silver epoxy. In a second step, above the free sample surface, cleavers were glued using silver epoxy. The \tmc single crystals were cleaved at room temperature inside the STM chamber and introduced into the cold STM head at a temperature of about 100~K (please see appendix \ref{sample_prep} for more details regarding the sample preparation and our efforts to minimize internal stress effects). This preparation procedure is reliable and was used by one of us on another very fragile needle-like inorganic quasi-1D material: NbSe$_3$\cite{Brun2009}. 

The \tmc samples were further cooled down to helium temperature. Between 45~K and 4.2~K the samples were slowly cooled down to maintain a cooling speed of about 1.5~K per hour in order to ensure a very good structural ordering of the ClO$_4$ ions and favor the bulk development of the superconducting phase. The STM/STS experiments were carried out at a base temperature of 300~mK using a homemade apparatus based on a $^3$He single-shot cryostat \cite{Serrier2013,Brun2014,Brun2017}. Mechanically cut PtIr tips were used. The differential tunneling conductance $dI/dV$ spectra were obtained from the numerical derivative of single $I(V)$ curves. The individual spectra were further convolved with a Gaussian filter compatible with the thermal broadening at 300~mK.


\section{Results}
Figure~\ref{fig1_STM_topos} shows the result of a nicely cleaved \tmc sample. Panel a) shows a $400\times330~\mathrm{nm}^2$ STM topography measured at $T=300$~mK emphasizing several (\textbf{a},\textbf{b}) crystallographic planes separated by elementary steps oriented along the \textbf{c} direction. This is corroborated by the $Z(Y)$ profile measured along the vertical dashed line and presented in panel b). This profile shows flat sections of molecular (\textbf{a},\textbf{b}) planes separated by an average $c$ parameter of about $c\simeq$1.49~nm as measured by STM locally in this particular area. It is consistent with the bulk reported value of 1.33~nm but presents an increase of about ten percent. This is a general observation that we made on other areas or samples. Additionally, on  rare occasions, we have been able to obtain a clear molecular resolution on the (\textbf{a},\textbf{b}) planes, as shown in panel c). Here, we clearly see the quasi-1D chains lying along \textbf{a}. Note that a tip change occurs in the bottom part of the image. Our results emphasize for the first time that the electronic properties of \tmc samples are accessible to STM/STS measurements if care is taken in the sample mounting and cleavage process. Based on \cite{Alemany2014} by Alemany et al., we expect that the LDOS as measured by STM should be dominated by the HOMO orbitals of the TMTSF molecules, with almost no weight on the ClO$_4$ ions.

Taking advantage of being able to scan crystallographic surface planes, we investigated the low-energy electronic properties of \tmc at various locations. The bulk critical temperature $T_c\simeq1.2$~K was precisely determined from \textit{ex-situ} electronic transport measurements on samples taken from the same batch, revealing a critical field $H_{C_2-c^*}\approx$0.2~T along \textbf{c$^*$}, in good agreement with the literature (see appendix~\ref{elec_resistivity}). Overall, our experimental results carried out on four different samples from the same batch all revealed inhomogeneous electronic properties at the surface, as measured by the STM/STS technique. This is in strong contrast to what can be expected from the $T_c$ and bulk homogeneousness of the electronic properties as probed using macroscopic techniques for a slow sample cooling speed \cite{Jerome16,Yonezawa2018}.

\begin{figure}[tbh]\begin{center}\includegraphics[width=\linewidth]{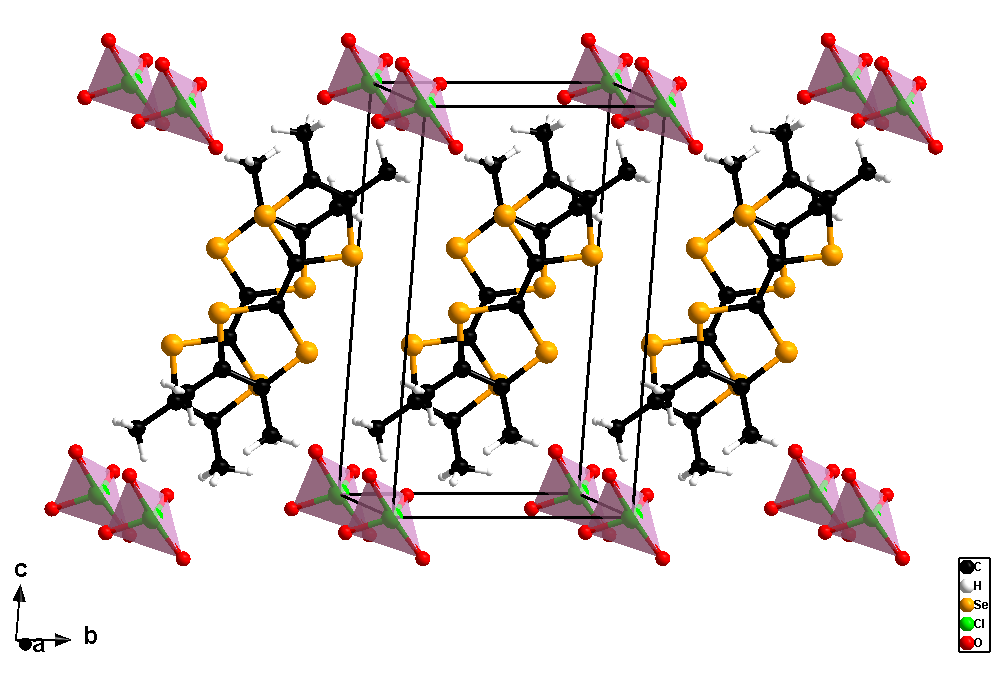}\caption{(color online) Crystal structure of \tmc viewed along the \textbf{a} direction. Note that the final ClO$_4$ anions order in the crystal structure depends on the cooling procedure: they can be either disordered or ordered with a doubling of the \textbf{b} period (see text).
\label{structure_a}}\end{center}\end{figure}

\begin{figure}[tbh]\begin{center}\includegraphics[width=\linewidth]{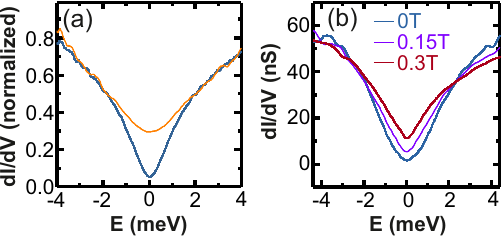}\caption{(color online) Tunneling spectroscopic properties of \tmc cleaved under ultrahigh vacuum and acquired by scanning tunneling spectroscopy in the area shown on the lower terrace of the figure~\ref{fig1_STM_topos}a). Differential tunneling conductance $dI/dV(E=eV_{bias})$ spectra measured in the energy range [-4;+4]~mV with the following set-point for STS: $I=300$~pA and $V_{bias}$=-9~mV. Each presented spectrum is the average of a $60\times60$ STS grid measured at $T=300$~mK (blue spectrum) or $T=2.13$~K (yellow spectrum) over the same area of $8\times8$~nm$^2$. b) Magnetic field dependence of the $dI/dV(E=eV_{bias})$ spectra at $T=300$~mK with B applied along \textbf{c$^*$} between 0 and 0.3~T (blue: 0~T, purple: 0.15~T, red: 0.3~T). The magnetic field dependence was probed in another sample area than the one probed in panel a). \label{fig2_STS_spec}}\end{center}\end{figure}

In the present work, we ascribe the electronic inhomogeneities observed at the surface to a structural disorder induced in the anions sublattice. Indeed the ClO$_4$ ions are situated in-between molecular (\textbf{a},\textbf{b}) planes, as can be seen in the figure~\ref{structure_a}. As the natural cleavage plane parallel to (\textbf{a},\textbf{b}) is likely to pass in between the molecular (\textbf{a},\textbf{b}) planes, i.e. through the ClO$_4$ ions, the cleavage process is expected to statistically remove a large part of these anions. Such an effect was already reported earlier by one of us in another well-known quasi-1D inorganic material, called the blue bronze, developping a charge-density wave ground state where the ions are located in-between the molybdenum oxydes planes \cite{Brun2005,MachadoCharry2006}. In \tmc samples cleaved under ultrahigh vacuum we could identify at $T=300$~mK different phases at the surface, including superconducting regions and SDW areas. In addition, other regions presenting correlated metallic states were also observed. We present below the characteristic tunneling spectra measured for each type of region.

Figure~\ref{fig2_STS_spec} shows differential tunneling conductance measurements performed both below $T_c$ in the superconducting state and above $T_c$. The panel ~\ref{fig2_STS_spec}a) shows representative $dI/dV(E=eV_{\mathrm{bias}})$ spectra measured at $T=300$~mK (blue spectrum) and at $T=2.13$~K (yellow spectrum), in zero-field cooled sample, in a $8\times8$~nm$^2$ region located in the lower terrace of the sample area seen in Fig.\ref{fig1_STM_topos}a)  indicated by a white square. The comparison between these two spectra enables to associate the internal dip structure of the $300$~mK spectrum to the superconducting local density-of-states (LDOS). This is confirmed by the panel ~\ref{fig2_STS_spec}b) presenting the magnetic field dependence of the $dI/dV$ spectra at $T=300$~mK for a magnetic field applied along \textbf{c$^*$} up to 0.3~T. A continuous increase of the conductance around $E_F$ is observed with increasing magnetic field. A maximum of conductance is achieved around 0.25-0.3~T that we attribute to the surface critical field along \textbf{c$^*$} $H_{C2S}$. This is in good quantitative agreement with bulk $H_{C2}\approx0.2~T$ measurements \cite{Yonezawa12} as well as our own electrical resistivity measurements along \textbf{c$^*$} (see appendix~\ref{elec_resistivity}). We provide in the appendix other $dI/dV$ spectra showing similar superconducting characteristics but emphasizing various gap filling, measured on other areas or samples (see appendix~\ref{add_sc_measurements}).

The figure~\ref{fig3_STS_spec} presents a typical tunneling spectrum, acquired in the same superconducting region as the low-energy spectra presented in panel~\ref{fig2_STS_spec}a), but over a much larger energy range of about 50 times the superconducting energy gap. It is seen that the characteristic spectrum is strongly V-shaped up to more than 40~meV. 

\begin{figure}[tbh]\begin{center}\includegraphics[width=\linewidth]{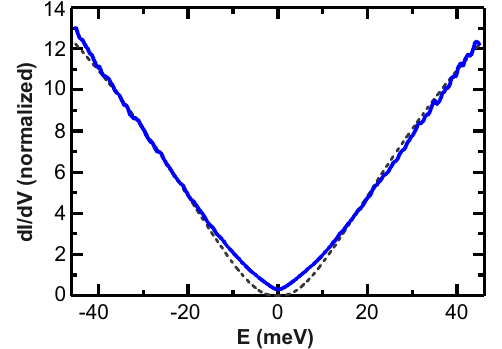}\caption{(color online) Tunneling spectroscopic properties of \tmc acquired by scanning tunneling spectroscopy in the area shown on the lower terrace of the figure~\ref{fig1_STM_topos}a), where the superconducting spectra shown in Fig.~\ref{fig2_STS_spec}a) were also acquired. Differential tunneling conductance $dI/dV(E=eV_{bias})$ spectra measured in the energy range [-40;+40]~meV with the set-point: $I=300$~pA and $V_{bias}$=-40~mV. The presented spectrum (continuous thicker curve) is the average of a $60\times60$ STS grid measured at $T=2.13$~K over a $8\times8$~nm$^2$ area. The same normalization procedure as in Fig.~\ref{fig2_STS_spec}a) is applied. Comparison with the Anderson-Hubbard model (dashed thinner curve), modelling the influence of disorder on the DOS at high energy, see text. \label{fig3_STS_spec}}\end{center}\end{figure}


\begin{figure}[tbh]\begin{center}\includegraphics[width=\linewidth]{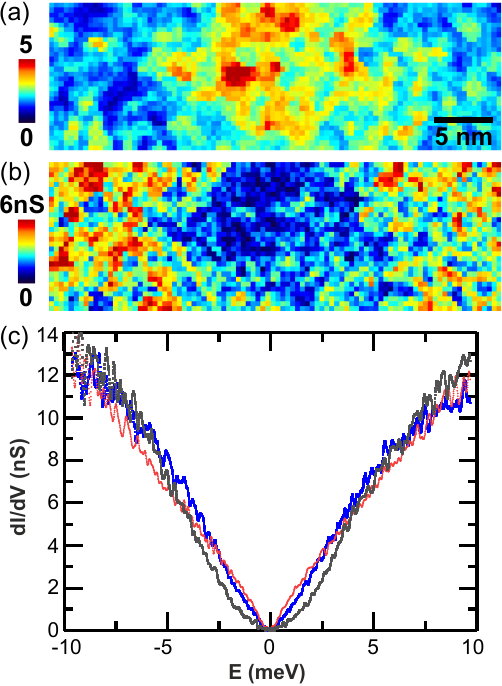}\caption{(color online) Measurements done at $T=300$~mK. a) Spatial map presenting the exponent $\eta$ fitted from individual $dI/dV(E)$ spectra assuming the energy dependence $dI/dV(E)\propto E^{\eta}$ acquired on a 92$\times$29 grid on a 40$\times$13~nm$^2$ area. The scale bar is indicated. b) Differential tunneling conductance $dI/dV(E)$ map plotted at -2~meV, measured over the same 13$\times$40~nm$^2$ area, extracted from a full $I(V)$ grid in the energy range [-10;+10]~mV with the set-point for STS $I=300$~pA and $V_{bias}$=-9~mV. Three different neighboring regions are seen: left superconducting (red-yellow-green/brighter), middle spin-density-wave (blue/darker region), right superconducting (red-yellow-green/brighter). c) Representative $dI/dV(E)$ spectra averaged locally in each left, middle, right area of the conductance map presented in panel b). Each spectrum is the average of about 50 single spectra. \label{fig3_V_U_spec}}\end{center}\end{figure}

Figure~\ref{fig3_V_U_spec} presents an interesting situation of local inhomogeneous electronic properties. It features a 40$\times$13~nm$^2$ region where a spin-density wave puddle of size 15-20~nm (blue/darker area in panel b) is comprised between two superconducting areas (red-yellow-green/brighter area in panel b). The local tunneling spectra (blue and red/lighter colors in panel c) corresponding to the superconducting areas are similar to the one presented in figure~\ref{fig2_STS_spec}a) but reach zero conductance very close to zero energy. In contrast, the middle area presents a strongly U-shaped spectrum (black curve) with a concavity that is opposite to the one of the superconducting spectra. Moreover, this U-shaped spectrum presents a larger energy gap and a lower conductance around the Fermi level reaching zero at $E_F$. In order to further support this result, we heuristically fitted all $dI/dV$ spectra acquired over a 92$\times$29 grid in this area, assuming a power law energy dependence in the form $dI/dV(E)\propto E^{\eta}$. The extracted exponent map $\eta(x,y)$ is presented in panel a). It anti-correlates with the conductance map of panel b): The larger the local exponent $\eta(x,y)$ (SDW puddle) the smaller the local conductance around -2~meV. Our result is in line with the reported SDW energy gap of about 3~meV \cite{Bando85}, 4.6~meV \cite{ICHIMURA2001}, up to 6~meV \cite{Fournel83}. Further analysis of the experimental local exponent variations shown in panel~\ref{fig3_V_U_spec}a) enables to extract a superconducting and SDW coherence length of about 5~nm along the chain direction, implying a low mean-free path of about 1~nm. As we will see below, these low values with respect to the bulk clean limit ones, are consistent with the high broadening parameter required to fit the single-particle spectra. We have found numerous SDW puddles at the surface, additional examples being presented in the appendix~\ref{add_sdw_measurements}. 

In addition to the superconducting and SDW puddles we have also found local regions where the tunneling spectra present only a correlated V-shaped background at large energy but without a clear signature of superconducting dip or SDW gap at low energy. Such regions could correspond to puddles of weakened or destroyed superconductivity having additional in-gap filled states,  as for instance proposed in a variable-cooling transport study to describe a granular superconducting network \cite{Yonezawa2018}. Examples of this case are shown in appendix~\ref{add_high_energy_measurements}. Finally, we have also encountered several locally insulating planes at the surface, where it was not possible to perform STS of the low-energy spectrum. In such cases, the STM tip dug several nanometers or tens of nanometers into the interior of the material in order to find low-energy conducting planes satisfying a set-point such as $I=200$ to $300$~pA and $V_{bias}$=-5 to -9~mV. We interpret such situations as damaged/defected surface parts of the sample where some regions of the surface crystallographic planes are electrically badly connected to bulk parts preventing the establishment of a tunneling current at low voltages. An example of such a situation is provided in the appendix~\ref{tip_digging}, where the area could be imaged using a larger bias voltage before and after the tip dug into the surface while performing low-energy STS.

We summarize as follows the general observations that could be inferred from our numerous investigations : \newline i) Superconducting regions are fewer at the surface plane than expected in the bulk. \newline ii) Superconducting coherence peaks are strongly smeared out. \newline iii) In the superconducting regions, the zero-bias conductance i.e. the LDOS at the Fermi energy, is most often finite and strongly fluctuates from one location to another. \newline iv)  In the SDW regions, the single-particle excitation spectra are U-shaped and reach zero at the Fermi energy. \newline v) The normal state presents a strong V-shape background extending to energies at least two decades above the superconducting energy gap.

\section{Theory and Discussion}

From electronic bandstructure calculations, a strongly V-shaped LDOS is not expected around $E_F$ \cite{Alemany2014}. The existence of a quasi-1D bandstructure could lead to a $k_{\parallel}$ tunneling dependence but not as strong as to induce such a fast increase of conductance as observed in our case over only $40$~meV assuming a reasonable sample work function of at least one eV. A more convincing explanation is that we measure an intrinsic V-shaped LDOS due to the strongly correlated quasi-1D character of the LDOS combined to a rather large surface disorder associated to an inhomogeneous granular state. This emerges by first noting that the pronounced background depletion of LDOS seen up to at least 40~meV in Fig.~\ref{fig3_STS_spec} is not uncommon in disordered materials \cite{McMillan80,Hertel83,Carbillet20,Richardella10}, including layered organic superconductors \cite{Diehl15,Guterding16}. These are characterized by sizable  electron-electron interactions that can couple to disorder and then alter the density of states over relatively large energy scales. The perturbative approach of Altshuler and Aronov to this form of coupling is well known to produce such a suppression of the tunneling density of states  \cite{Altshuler80}. Analogous features can also be found from a numerical treatement of the  disordered Anderson-Hubbard model \cite{Shinaoka09b,Shinaoka09}. These conditions are likely to be fulfilled for a compound as (TMTSF)$_2$ClO$_4$ in  which quantum criticality  linking  superconductivity to a correlated antiferromagnetic insulating state is occuring in  the presence of non magnetic defects of the anion ordering.  As shown below, the coupling of electron-electron interaction to randomness is responsible for the shaping of LDOS in energy down to 10~meV or so. 

 We follow the numerical approach of Ref.~\cite{Shinaoka09b} to the Anderson-Hubbard model with  the expression for the DOS in two spatial dimensions, 
\begin{equation}
    N_{\mathrm{AH}}(E) = c \exp\big[ -\alpha \ln^2\big(\sqrt{2}|E|/E_F\big)\big], 
    \label{DOSAH}
\end{equation} 
 where $c$ and $\alpha$ are positive constants and $E_F$ is the Fermi energy which is typically $E_F\simeq 0.26$~eV for a quasi-1D metal like (TMTSF)$_2$ClO$_4$ \cite{Jerome24}. Using a value of $\alpha\simeq 0.3$,  the above expression fits fairly well the data of Fig.~\ref{fig3_STS_spec} over the whole range of intermediate energy. The value of $\alpha$ is comparable to the one  found in layered organic superconductors showing a similar DOS depletion at intermediate energy \cite{Diehl15}. However, the data do not evolve to the formation of a soft Hubbard gap as  predicted to take place by the model around the Fermi level when the disorder-free electron system  is  a  correlated insulator with a sharp gap. Deviations found below $E_0\sim 10$~meV come from the onset of coherent quasi-particle  states with a finite DOS on which  our model developed below is based.


 Below $E_0$, the measured energy dependence of the STS spectra of Figures~\ref{fig2_STS_spec} and \ref{fig3_V_U_spec}, associated with superconducting regions, is linear-like all the way down to the energy gap region. There, the LDOS is most often finite at zero $E$ as in fig~\ref{fig2_STS_spec} (see also fig.~\ref{fig_sts_sc_spectra} of the appendix~\ref{add_sc_measurements}) which is in agreement with macroscopic specific heat measurements \cite{Yonezawa12,Yonezawa25}, but can also reach zero locally as in fig~\ref{fig3_V_U_spec}. This is not consistent with an $s$-wave nodeless order parameter but reminds what is observed in $d$-wave cuprates superconductors by STS \cite{Fischer2007}. Furthermore, non-magnetic disorder in the ClO$_4$ anions lattice can be expected to create local defects, as a result of the cleaving process. A more natural explanation compatible with point group symmetry would then be a $d$-wave symmetry where line nodes would give a linear energy dependent DOS in the superconducting energy gap, consistent with the angular and the temperature dependence of the specific heat \cite{Yonezawa12,Yonezawa25}. Moreover for a $d$-wave symmetry non-magnetic point defects behave as pair breakers, leading to additional gap filling, as  predicted for a $d$-wave superconductor with impurity scattering in the unitary limit \cite{Maki98,Suzumura89} and  often observed experimentally. 


 In the interpretation of  the  whole low energy part of the STS spectrum below $E_0$, we shall adopt the view that  SC  and  SDW  states of the Bechgaard salts are the  result of  repulsive interactions alone, as approached  by  the theory of quasi-1D electron gas model.  This model is based upon an electron spectrum with strongly anisotropic near- and next to near- neighbor transfer integrals. These lead  to an open Fermi surface on which a set of electron-electron scattering amplitudes or coupling constants $g_i$ can be defined.  We shall make use of the  extensive studies of renormalization group (RG) method  for this model \cite{Bourbon09,Sedeki12,Nickel06} (See also appendix~\ref{model_Q1D_DOS}). At the one-loop level for instance, these give  a very reasonable description of the actual sequence of instabilities of the metallic state  against SDW and SC (\textit{d}-wave) orders as the  antinesting namely, the   interchain next-to-nearest neighbor  hopping  parameter, $t_\perp'$,  is tuned to simulate the influence of   pressure. This occurs  together with a dome of SDW quantum critical fluctuations that permeate deeply the metallic phase and whose amplitude  scales with the one  of $T_c$ under pressure. 
 
\begin{figure}[tbh]\begin{center}\includegraphics[width=\linewidth]{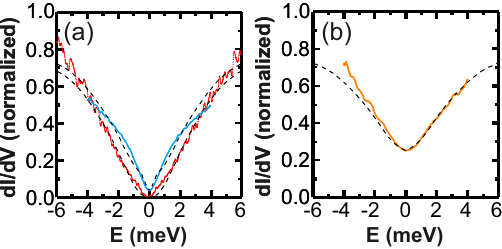}\caption{Comparison of the renormalization group prediction (dashed dotted lines) and STM experiments for the density of states at (a): $T=300$~mK in the SC state for the two regions shown in Figures \ref{fig2_STS_spec} and \ref{fig3_V_U_spec}, and (b): $T=$2~K in the metallic state in the region shown in Figure \ref{fig2_STS_spec}-(a). \label{STMRG}}\end{center}\end{figure}

 At the two-loop level of the RG, the one-electron spectral properties of the quantum critical domain of the model can be obtained. This the case of the energy profile of the  DOS (normalized by the bare part $N_0$), $ 
 \bar{N}_n(E) = - \frac{2}{ N_0 \pi} \frac{1}{V} \sum_{{\bf k}} {\rm Im}\, G({\bf k},E)$, which is  linked to the coherent part of  the one-particle retarded Green function $G$, as a function of wave vector ${\bf k} $ and energy distance $E$ from the Fermi level.  Its renormalization is governed by the diagrammatic RG equation $\mathcal{O}(g_i^2)$ for the self-energy \cite{Sedeki12}, as portrayed in Fig.~\ref{Self} of appendix~\ref{model_Q1D_DOS}.  The local  DOS probed by single particle tunneling, $\bar{N}_n(E) = \langle z({k_\perp}, E)\rangle_{k_\perp}$, is  expressed in terms of the  energy dependent quasi-particle weight $z$ averaged over the transverse wavevector $k_\perp$. In the metallic phase  for $T>T_{c,\mathrm{SDW} }$ calculations predict a drop of $\bar{N}_n(E)$  upon decreasing $|E|$ that is well described by a power law
 \begin{equation}
     \bar{N}_n(E) \propto \big(\mathrm{Max}\{\pi k_BT,|E|\}\big)^\eta, \ \ \ E< 2t_\perp'
\label{Nn}
 \end{equation}
 for low energy up to the  scale of antinesting  $2t_\perp'$ of the order of 5~meV for a quasi-1D compound like (TMTSF)$_2$ClO$_4$ (see appendix~\ref{model_Q1D_DOS}). This scale marks the onset of an interplay between SDW and \textit{d}-wave pairing fluctuations that initiates the low energy  quantum critical domain. At higher energy for  $E> 2t_\perp'$, the calculations show a transient  weaker energy dependence of $\bar{N}_n(E)$ (See for instance Fig.~\ref{STMRG}). The latter  evolves towards the higher energy scale of the nearest neighbor interchain hopping $t_\perp\sim 10t_\perp'\sim E_0 $, located in turn at the onset of randomness scale for the DOS.  In the low energy sector of interest, the exponent $\eta(t_\perp')$ is non universal being  $t_\perp'$ or pressure dependent. Close to the QCP where SC emerges on the brink of SDW, $t_\perp'\simeq t_\perp'^*$ and  $\eta(t_\perp'^*) \simeq 1$ with  $\bar{N}_n(E)$ exhibiting a linear energy profile.  On the SDW side of the phase diagram, at lower  $t_\perp'< t_\perp'^*$, $\eta(t_\perp')>1$  increases above unity and $\bar{N}_n(E)$ shows a concave U-shape behavior  in energy for temperature $T>T_{\mathrm{SDW}}$ in the metallic state. By contrast,  moving away from the QCP on the SC side, $\eta(t_\perp')$  decays below unity, a drop that correlates with the one of $T_c$ with $t_\perp'$. The influence of non magnetic impurity scattering on $\bar{N}_n(E)$ is here neglected since its impact on the strength of electron correlations of the  quantum critical domain can be considered  to be small \cite{Sedeki18}.
 
At sufficiently low $|E|$, the metallic $\bar{N}_n(E)$  will be connected   to the suppression of the DOS following the opening of a  gap $\Delta_\mu$ from either $\mu=\mathrm{SC}$ or SDW long-range order. The normalized  DOS will take the form
\begin{equation}
\label{Ns}
\bar{N}(E)= \bar{N}_n(\Delta_\mu,|E|)\Big\langle{\rm Re}\Bigg\{ {|E+ i\Gamma| \over \sqrt{ (E+ i\Gamma)^2 - \Delta_\mu^2(k_\perp)}}\Bigg\} \Big\rangle_{k_\perp},
\end{equation}
where ${\bar{N}_n(\Delta_\mu,|E|) \propto (\text{Max}\{\Delta_\mu,|E|\})^\eta }$.  The $k_\perp$-average  in brackets is a normalized BCS expression for the DOS of the quasi-1D electron gas in  the presence of a gap $\Delta_\mu$ and the  broadening $\Gamma$ due to impurity scattering  assumed to be pair-breaking for both  orderings \cite{Dynes78,Suzumura89}.  

 The above expression for the DOS enters in the differential tunneling conductance at temperature $T$:
   \begin{equation}
\label{dIdVCal}
{dI\over dV}(E) =   G_{nn}\int_{-\infty}^{+\infty} \bar{N}(\omega)\big[ -{\partial f(\omega+ eV)\over \partial (eV)}\big] d\omega + G_0,
\end{equation} 
where $E=eV$, $f(\omega)$ is the Fermi distribution, $G_{nn}$ and $G_0$  are adjustable  constants for the  normal conductance and the finite  LDOS at the Fermi level.    
  
To fit the above expressions with the low energy STM data at $T=300$~mK in  the SC state, one substitutes the $k_\perp$ - dependent, \textit{d}-wave, superconducting gap  $\Delta_{\mathrm{SC}}(k_\perp) = \Delta_{\mathrm{SC}} \cos k_\perp$ for an open quasi-1D Fermi surface, together with the RG results for $N_n(E)$  obtained from typical  figures for the couplings $g_i$ and   band parameters for the Bechgaard salts \cite{Sedeki12} (See appendix~\ref{model_Q1D_DOS}). These lead to a calculated critical temperature  $T_c\simeq 1.1$~K at the value of antinesting pressure parameter $t_\perp'/k_B=26$~K, namely close to the QCP, a  value of the magnitude found experimentally in the bulk. The impact of impurity scattering on the ordering will be incorporated through  the broadening factor $\Gamma$. The calculated energy profile of $dI/dV$    is shown   in Fig.~\ref{STMRG}-(a) with the STM data at $T=300$~mK for the data of two different SC areas shown in Figures \ref{fig2_STS_spec} and \ref{fig3_V_U_spec}. The best fit is  obtained for ${\Gamma/\Delta_{\mathrm{SC}}= 0.6}$, $\Delta_{\mathrm{SC}}\simeq$0.76~meV, and a ratio $\Delta_{\mathrm{SC}}/k_BT_c=8$ that exceeds by a factor about four the BCS value for the SC state, a ratio  known to be partly enhanced by impurity scattering \cite{Suzumura89}. The prediction gives a fair description of  the differential conductance  extending from the  Fermi level up to $|E|\sim 5$~meV ($\gg\Delta_{\mathrm{SC}}$). At $|E| <\Delta_{\mathrm{SC}}$, this is consistent  with  a \textit{d}-wave gap having nodes on the Fermi surface which  leads to a linear dependence of the DOS on energy. 
A sizable broadening factor $\Gamma$ accounting  for the  enhanced impurity scattering  at the surface  is responsible for the suppression of the coherence peaks. This behavior then  joins  the  $E$-linear  DOS  caused by quantum fluctuations spreading over  a sizeable energy interval well above the gap in accord with the RG solution close to the QCP.   

In  the normal phase above $ T_c$,   the     ${T=2}$~K data of the differential conductance of Fig.~\ref{fig2_STS_spec}-(a) and reproduced  in Fig.~\ref{STMRG}-(b)  show  a significant filling of the DOS  expected by   temperature effects close to the Fermi level.     In that case Eq.~(\ref{dIdVCal}) reduces to the normal state expression $(dI/dV)(E)\propto \bar{N}_n(E)$, where $\bar{N}_n(E)$ is given by the  RG results (\ref{Nn}) computed at $T=2$~K with the same   parameters that led to the $T_c$  discussed above \cite{Sedeki12} (see appendix~\ref{model_Q1D_DOS}).  $\bar{N}_n(E)$ is $E$-linear ($\eta\simeq 1$)  outside the thermal broadening region at $|E| > \pi k_BT$   up to the beginning of the  weaker energy transient dependence taking place from   5 meV or so. This behavior is congruent with the data of Fig.~\ref{STMRG}-(b)  obtained up to 4 meV.

\begin{figure}[tbh]\begin{center}\includegraphics[width=\linewidth]{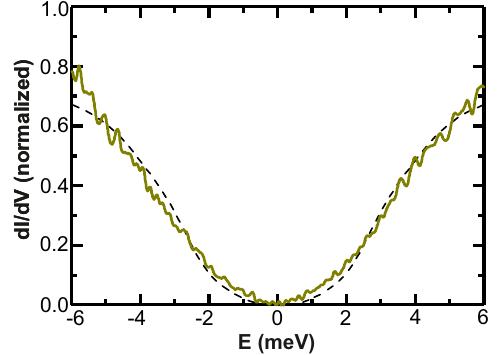}\caption{Comparison of the normal state renormalization group prediction (dashed line) and STM experiments of Fig.~\ref{fig3_V_U_spec} for the density of states in the SDW state at  $T=300$~mK. The theoretical prediction including in addition the SDW gap in the DOS is shown in figure~\ref{conducRG} of appendix~\ref{model_tunnel_cond_ord_state}.\label{STMRGSDW}}\end{center}\end{figure}

We now consider  the results of  the more  disordered region where  the system  turns to be   a SDW insulator at $T=300$~mK, as displayed in  Fig.~\ref{fig3_V_U_spec} and reproduced in Fig.~\ref{STMRGSDW}. In the phase diagram this  would correspond to a negative shift along the pressure axis towards the SDW region. According to the figures, the data show a noticeable concavity or  a U-shape form in energy for the DOS. This can be  simulated theoretically by a downward shift of antinesting at $t_\perp'/k_B\simeq 25.4 $K in the phase diagram of the RG solution. An instability of the metallic state against  SDW order then occurs at the critical temperature  $T_{\mathrm{SDW}}\simeq 5$~K,  a magnitude    typically  found in  quenched ClO$_4$ samples \cite{Tomic82,Schwenk84}.   In this SDW regime, the single particle gap $\Delta_{\mathrm{SDW}}(k_\perp)\approx \Delta_{\mathrm{SDW}} $  presents no nodes and  can be taken as  essentially momentum independent along the Fermi surface. From the normal state fit in  Fig.~\ref{STMRGSDW} obtained using  $\Delta_{\mathrm{SDW}}/k_BT_{\mathrm{SDW}}= 3.6$ and  $\Gamma/\Delta_{\mathrm{SDW}}= 0.6 $ in Eqs. (\ref{Ns}-\ref{dIdVCal}),  no coherence peaks for the DOS at $|E|\sim \Delta_{\mathrm{SDW}}\simeq1.55$~meV are found but a small kink is visible between $\pm$~2 and $\pm$~3 meV associated with a change in the energy dependence similar to the simulated curve shown in the bottom panel of figure\ref{conducRG} (see appendix~\ref{model_tunnel_cond_ord_state}) taking into account the smeared SDW gap. This is followed by a U-shape $E$-profile below the SDW gap scale, making scarcely visible the connection  with the normal  part (\ref{Nn}) characterized  by an exponent $\eta\simeq 1.3$ and a similar positive curvature in energy up to $2t_\perp'$.

It is worth noticing that the spatial map of the exponent $\eta$ displayed in Fig.~\ref{fig3_V_U_spec}(a), show experimental values between 0.75 for superconducting regions and 5 for SDW insulating ones. Such a range of $\eta$ values is compatible with the ones found theoretically when the RG solution is carried out at slightly weaker or stronger $t_\perp'/k_B$ around 26~K. When larger $\eta > 1$ values are encountered, stronger insulating  SDW gaps take place (see Figs.~\ref{NvsE}-\ref{eta} of appendix~\ref{model_tunnel_cond_ord_state}) which is also observed experimentally (see  Fig.~\ref{fig_sts_sdw_spectra} of appendix~\ref{add_sdw_measurements}). However, it cannot be excluded  that in these strongly insulating conditions the predictions of the disordered  Anderson-Hubbard model can lead to results comparable to those of $\bar{N}(E)$ in Eq.~(\ref{Ns}) for the DOS at low energy.

 
\section{Conclusion}
Since the discovery of  the Bechgaard salts as the first series of organic superconductors,   the nature of the superconducting order parameter in its prototype compound, (TMTSF)$_2$ClO$_4$, has long frame a major part of the debate surrounding superconductivity  in the  series. Revealing the intricate link of superconductivty with bordering magnetic spin-density wave and  quantum critical  metallic phases has become another determinant  issue regarding these materials.

The experimental and theoretical insights brought out by the present work have direct bearing upon these  matters. Of particular note is the linear-like or convex low energy dependence of the local excitation spectra that has been probed by scanning tunneling spectroscopy. The linear profile  appears to be akin to the Planckian linear temperature dependence of resistivity observed close to the QCP. The E-linear density of states is consistent with nodal d-wave  rather than nodeless s-wave symmetry for the order parameter in the superconducting phase. By contrast, the concave energy dependence of the local excitation spectra, found in more disordered surface conditions,  matches very well with the nodeless order parameter  expected in an insulating spin-density wave phase.

Our experimental results find a  quantitative support in the renormalization group approach to the quantum critical precursors of  both magnetic and superconducting orderings. This  manifests itself through the prediction of a power law  energy dependence $E^{\eta}$ for the density-of-states  at low-energy. The  exponent $\eta$ that characteristizes metallic quantum fluctuations varies in relation with the nature of order in the low temperature phase. For  superconductivity,  $\eta\lesssim 1$  whereas $1<\eta<5$ for  spin-density-wave, in  definite connection with the surface mapping of  tunneling conductance obtained in our experiments. 
This overall agreement between theory and experiments gives strong support for  instabilities in the organic \tmc crystals as being primarily mediated by electron-electron repulsion combined to quantum criticality, resulting in nodal d-wave superconductivity and spin-density wave phases. Both phases are found to coexist in different regions of  the surface, as a result of large anions disorder induced by the cleavage process.


\section
{Acknowledgements}
We thank Claude Pasquier, Nicolas Dupuis and Shingo Yonezawa for fruitful discussion. We would like to thank Magali Allain for helping us produce the images of the structure of \tmc.\newline

*christophe.brun@sorbonne-universite.fr, Claude.Bourbonnais@USherbrooke.ca

\appendix

\section{Electrical resistivity measurements}
\label{elec_resistivity}

\begin{figure*}[tbh]\begin{center}\includegraphics[width=1.1\textwidth]{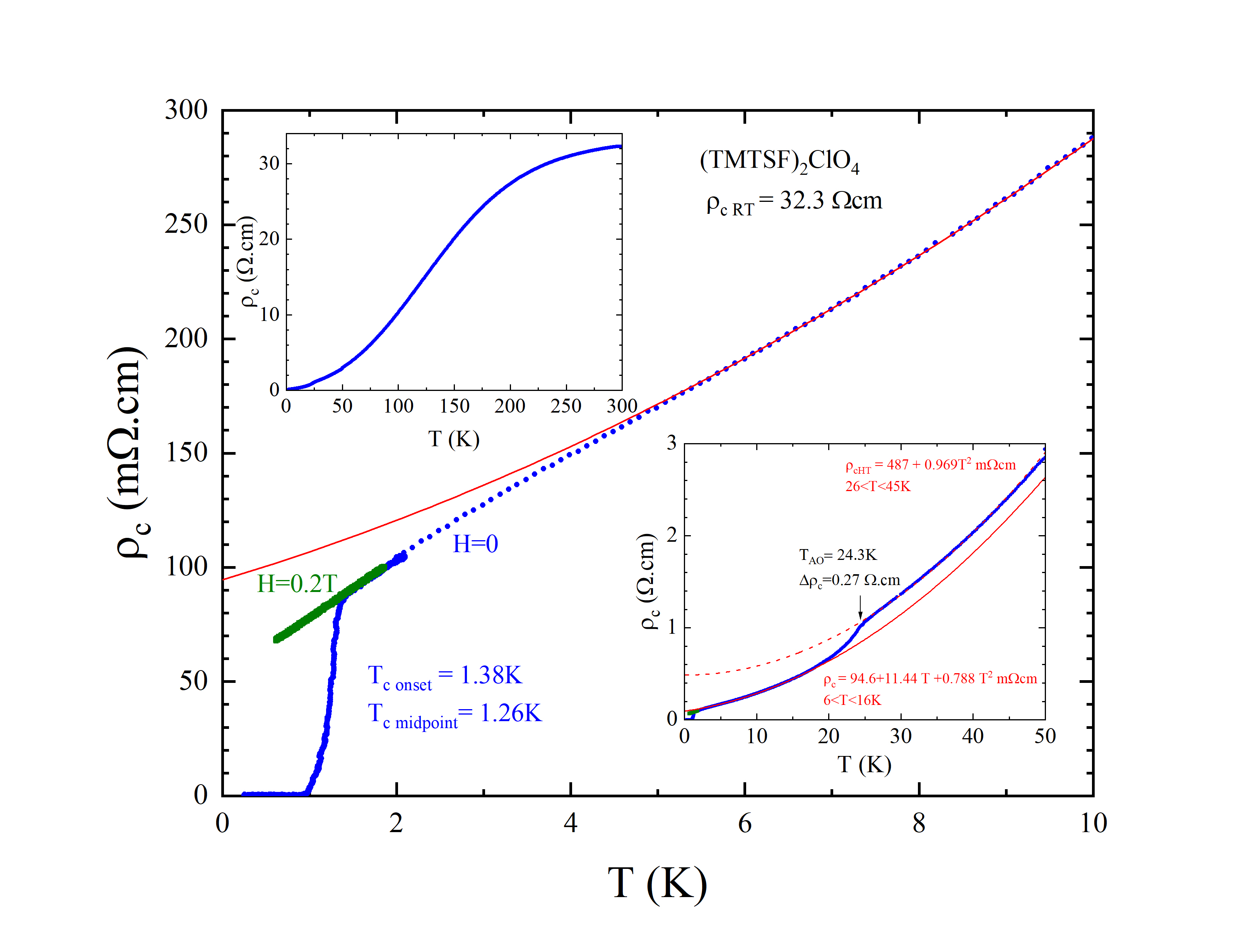}\caption{(color online) Temperature dependence of $\rho_c$, i.e. electrical resitivity along the c axis, for a very slowly cooled sample of $\mathrm{(TMTSF)_{2}ClO_{4}}$ measured between $T=300$~K and $300$~mK. The upper left inset displays the behaviour starting from $32.3\Omega.cm$ at room temperature. The lower right inset shows in more detail the behaviour below 50K with a drop in resistivity at 24.3K due to the ordering of $\mathrm{ClO_{4}}$ anions. Blue curves represent experimental measurements under zero magnetic field. The green curve represent experimental measurement under 0.2~T applied magnetic field the $c$ direction. The red curves (continuous and dashed lines) are fits according to the indicated formulas, see text. \label{fig_elec_resist}}\end{center}\end{figure*}

One $\mathrm{(TMTSF)_{2}ClO_{4}}$ single crystal from the same batch as the one used for STM/STS experiments, was selected for electrical transport measurements. The temperature dependence of the transverse resistivity $\rho_{c}$ has been measured in a Quantum Design physical property measurement system (PPMS) using an adiabatic demagnetization refrigerator adapted to the PPMS setup. The minimum temperature reached was 0.3~K at H=0 and 0.6~K at H=0.2~T. The resistance was measured applying a dc current of 50~µA. The cooling speed was decreased down to 0.01~K/min between 26 and 20~K in order to perform the ordering of the ClO$_4$ anions.

The resistivity shows a very large decrease when lowering the temperature between 300~K and about 2~K, as seen in the top left inset of figure~\ref{fig_elec_resist}. The overall behavior attests to very high sample quality, the residual resistivity ratio being larger than 340, in agreement with previous study \cite{Yonezawa2018}. The superconducting transition develops below 1.4~K with a midpoint $T_c$ of about 1.26~K. The critical magnetic field along the c direction is about 0.2~T (green curve).

As shown on Fig.~\ref{fig_elec_resist}, the inelastic part of the resistivity follows somewhat accurately around the anion ordering temperature up to about 50~K a polynomial law such as $\rho_{c}(T)$ $\sim AT + BT^2 $ displayed by the red dashed line on this figure. A temperature drop $\Delta\rho_c= 0.27 \Omega$.cm occurs between 25 and 20~K due to the anion ordering, see down right inset. However, the  experimental behaviour departs significantly from the polynomial law below 5K and is not affected by a magnetic field (see the green curve on the figure). Such a behaviour is suggestive of a paraconductive  contribution to the conductivity. The down turn of the resistivity below 5~K already reported in $\mathrm{(TMTSF)_{2}ClO_{4}}$ \cite{Auban11,Jerome24}, has actually  been  attributed to a SDW paraconductive contribution given by, 
$\Delta\sigma=\frac{\rho_{\rm n}-\rho_{\rm ex}}{\rho_{\rm n}\rho_{\rm ex}}$, where $\rho_{\rm ex}$ is the measured resistivity (blue dots on Fig.~\ref{fig_elec_resist}) while $\rho_{\rm n}$ is the behaviour of the inelastic  resistivity (red line) derived from fitting the  experimental  transport to a polynomial law  between 6 and 16~K . In addition we may notice the good quantitative agreement for the paraconductive contribution between the data  on Fig.~\ref{fig_elec_resist} and those reported in Ref.\cite{Jerome24}.

\section{crystal structure of \tmc}
\label{cryst_struct_app}
We provide here an additional view of the crystal structure unit cell of \tmc enabling seeing better the molecular stacking along the \textbf{a} direction.

\begin{figure}[tbh]\begin{center}\includegraphics[width=1.3\columnwidth]{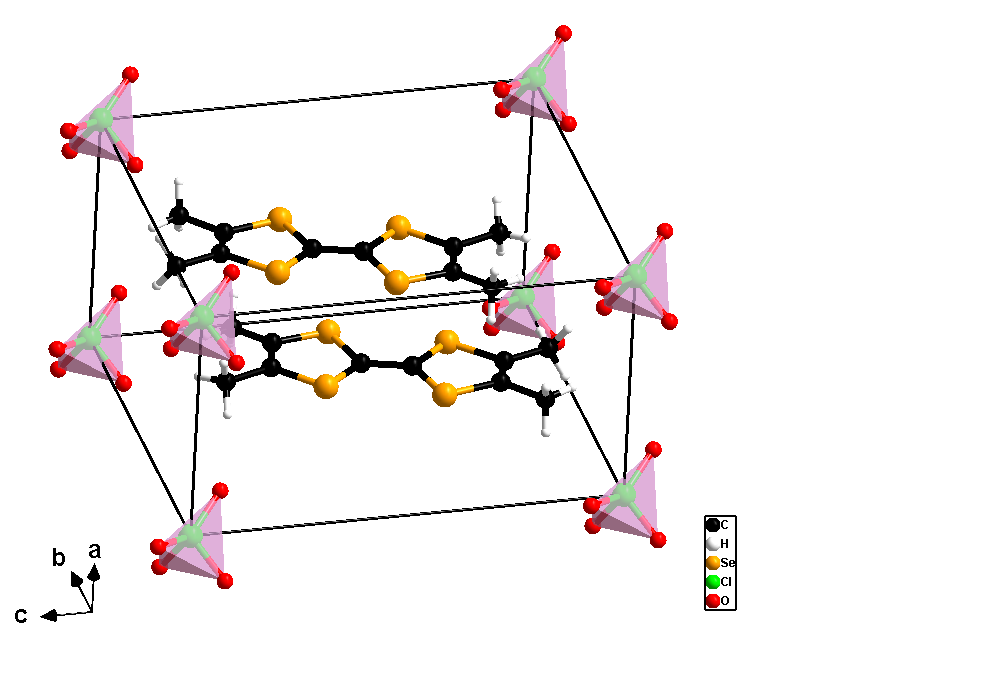}\caption{Unit cell of the crystal structure of \tmc.  \label{fig_cryst_struct_app}}\end{center}\end{figure}

\section{sample preparation-limiting internal stress}
\label{sample_prep}
\tmc single crystals are very fragile needle-like materials that are notoriously difficult to manipulate and prepare for STM/STS experiments. We provide here additional pieces of information about the sample preparation procedure we have followed in order to minimize internal stress effects induced by the mismatch of thermal expansion coefficients of the various materials used to hold the \tmc sample and cleave it. 

We took great care in gluing our samples. In particular in a first step, we glued the sample on a stainless steel sample holder using a silver epoxy polymerizing at room temperature to avoid any additional stress induced by sample heating and also to avoid chemical diffusion. In a second step, we tried to glue the cleaver by droping glue only on the sample surface, not clasping the edges. Then the cleavage process occurred \textbf{at room temperature} in the STM chamber before inserting the sample holder into the cold STM head. We believe that this way of doing might help avoiding clamping effect induced by different thermal coefficients of the glue, sample, sample holder and cleaver. In particular we verified optically in the STM chamber that the cleavage exposes a large part of the sample above the glue level along the \textbf{c*} direction. 

Thus, upon cooling, the part of the sample embedded deeply in the glue should indeed be subject to a large internal stress but is not the one that is measured. On the contrary the cleaved sample surface is connected to a sample part that is held well above the glue level along \textbf{c*}, enabling in principle the underlying stress in the bottom part of the sample to relax in the transverse direction. Nevertheless, despite all the care taken, an additional internal stress induced by the various thermal expansion coefficients of the different materials used could appear and induce additional disorder in the cleaved \tmc sample affecting in the end its surface electronic properties. This said, our experience with another very fragile inorganic quasi-1D needle-like compound NbSe$_3$ showed us that this preparation procedure in principle works and enables in this latter case to recover at the surface electronic properties similar to the bulk ones \cite{Brun2009}.

\section{sample cooling curve for STM/STS measurements}
\label{sample_cooling}

\begin{figure}[tbh]\begin{center}\includegraphics[width=\columnwidth]{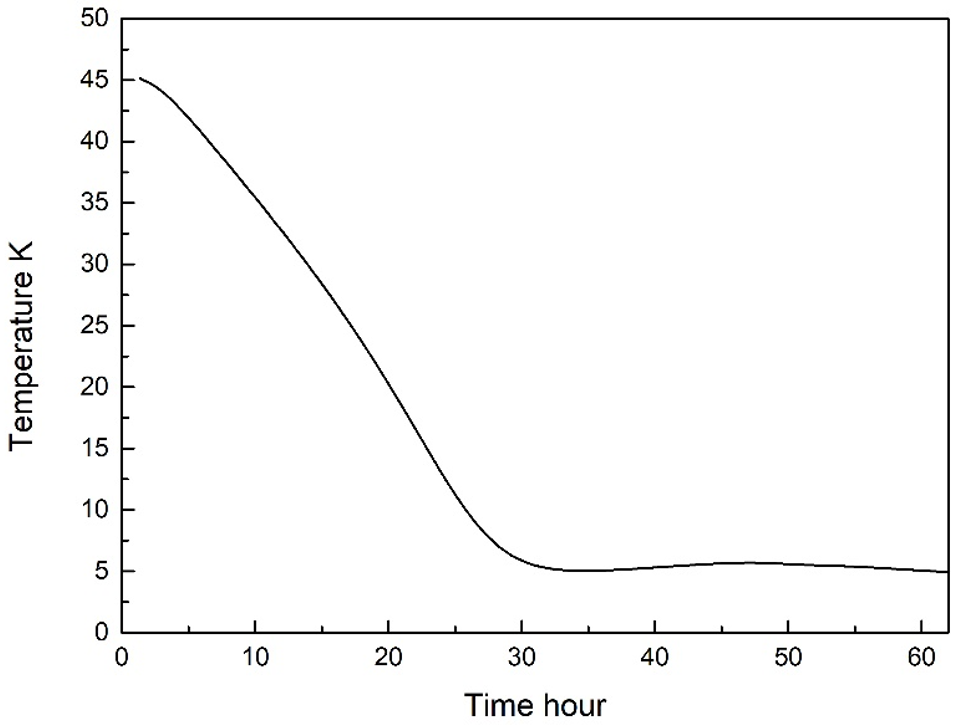}\caption{Cooling speed followed for all four (TMTSF)$_2$ClO$_4$ studied samples by STM/STS experiments in the present study.  \label{fig_stm_cooling}}\end{center}\end{figure}

Below 45~K, the (TMTSF)$_2$ClO$_4$ samples used in the present study were cooled down very slowly at cooling rates between 1 and 2~K per hour. This is presented in figure~\ref{fig_stm_cooling}. This ensures that the structural ordering of the anions is best realized in the material, leading to the largest superconducting bulk fraction below the superconducting critical temperature \cite{Jerome24}.

\section{additionnal STS measurements implying tip digging}
\label{tip_digging}

In nicely cleaved (TMTSF)$_2$ClO$_4$ areas, it was usually possible to scan the STM tip using constant current conditions with closed feedback loop using $I=10$ to 50~pA and $V=$-0.1~V. This corresponds to the case of the figure~1a) of the main text and to the figure~\ref{fig_stm_tip_dig} of the supplementary material. Indeed nice and wide molecular terraces are seen in both cases, separated by single or multiple lattice constant along the c axis. Nevertheless, it was not always possible to go from these scanning conditions corresponding to a 2-10~G$\Omega$ tunneling resistance to much lower tunneling resistance, i.e. about 10~M$\Omega$ needed to measure the superconducting excitation spectrum with enough signal to noise ratio. In some sample areas, when changing the scanning conditions to acquire a superconducting excitation spectrum, the STM tip dug several nm or even tens of nm below the originally flat sample surface. Such a behavior is exemplified in fig.~\ref{fig_stm_tip_dig}, the $Z(X)$ tip profile enabling to appreciate the depth (about 25~nm) and width (about 50~nm) of the hole created by the tip. In particular it can be noticed that the depth of this hole is about twice larger than the original tip height variation along this profile (about 10~nm) before the hole was made. 

\begin{figure*}[tbh]\begin{center}\includegraphics[width=\textwidth]{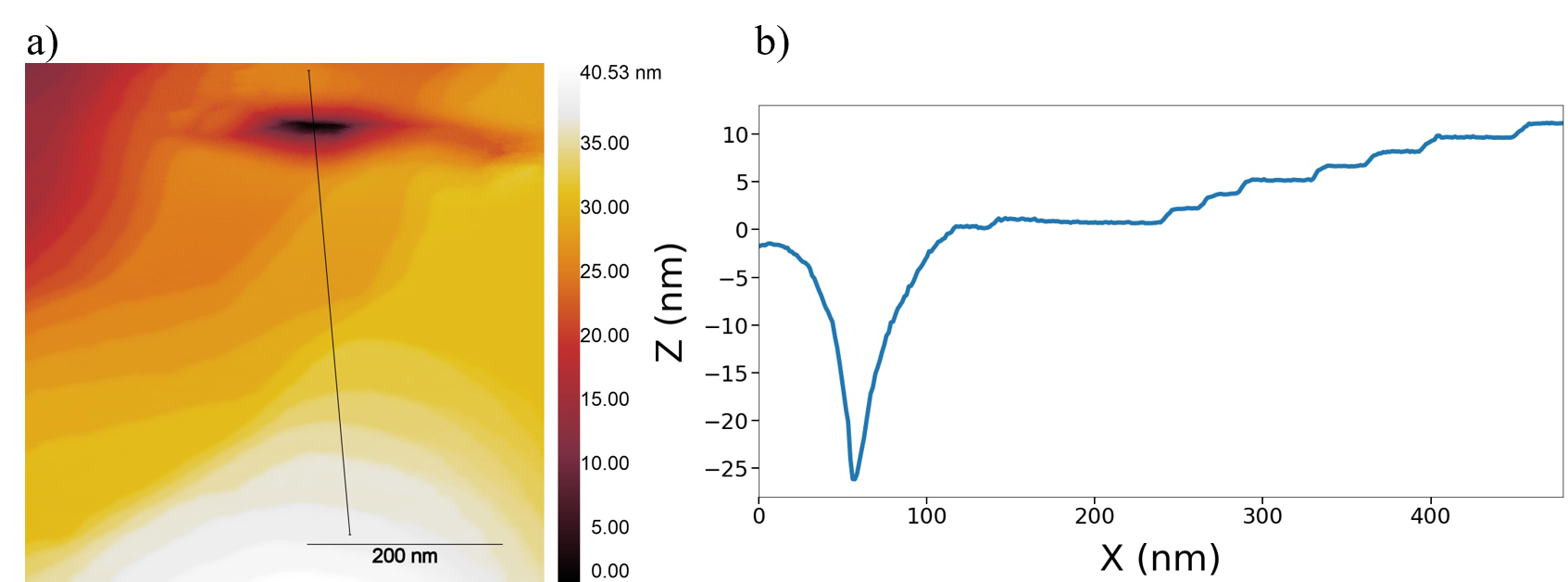}\caption{(color online) a) Raw color-coded constant current $Z(x,y)$ STM topography of a  (TMTSF)$_2$ClO$_4$ area of 550$\times$550~nm$^2$ measured with $I=50$~pA and $V=$-0.1~V. The color bar is shown on the right and varies between 0 and about 40~nm. Numerous molecular terraces are seen corresponding to flat areas of uniform color. The top black feature crossed by the continuous almost vertical line is a hole created by the STM tip into the surface after attempting measuring the superconducting excitation spectrum with a set-point $I=200$~pA and $V=$-5~mV. b) The STM tip profile $Z(X)$ measured in nm along the continuous black line shown in panel a) enables to quantify how deep the dip dug into the surface, i.e. about 25~nm, while the sample surface was originally a flat molecular plane at this location. In addition, this profile enables to see the regular spacing in z variations corresponding to single or multiple molecular planes of (TMTSF)$_2$ClO$_4$ along the \textbf{c} direction. \label{fig_stm_tip_dig}}\end{center}\end{figure*}

\section{additionnal STS measurements in superconducting regions}
\label{add_sc_measurements}

\begin{figure}[tbh]\begin{center}\includegraphics[width=\columnwidth]{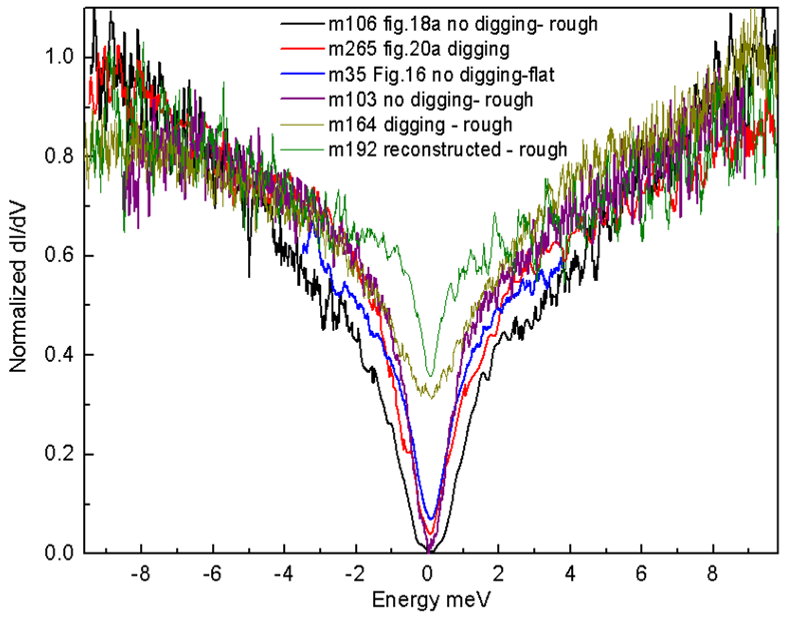}\caption{(color online) This figure summarizes representative superconducting excitation spectra measured at different locations and for different (TMTSF)$_2$ClO$_4$ samples from the same batch. For each spectrum it is indicated which experimental conditions were encountered in the corresponding location. In particular it is written whether the excitation spectrum could be measured at the top molecular surface plane (it is then written "no-digging") or not (it is then written "digging"). In addition, the term "rough" or "flat" is indicated to qualify the local surface roughness measured while scanning with typical conditions such as in figure \ref{fig_stm_tip_dig} (see text). The term "flat" means that the local tip height variations were below 0.1~nm (such as in fig.1c of the main text). The term "rough" means that the local tip height variations ranged from several angstroms to about 1~nm. The spectrum denoted "reconstructed" was measured in an area modified by the tip after it dug into the surface. \label{fig_sts_sc_spectra}}\end{center}\end{figure}

In order to be complete, we present here the entire diversity of single excitation spectra corresponding to superconducting regions measured on various areas and different (TMTSF)$_2$ClO$_4$ samples from the same batch. Figure~\ref{fig_sts_sc_spectra} presents these additional STS $dI/dV$ spectra. 

It is seen that the higher energy background, i.e. above $\approx$4meV is very similar among these different spectra. Nevertheless, the low-energy part, corresponding to the superconducting excitation spectrum, varies considerably. In particular the gap filling varies considerably from ones reaching zero exactly around the Fermi level (black spectrum in Fig.~\ref{fig_sts_sc_spectra} or in figure~\ref{fig3_V_U_spec} from the main text), very large gap filling like the m164 or m192 spectra. However, these spectra also share an almost linear low-energy behavior, in agreement with the other representative spectra presented in the main text. As the spectra presented in the main text, the superconducting coherence peaks are smeared out due to disorder, the presence of a strong V-shaped background and temperature broadening. 

In addition, we also presented in this figure the various cases encountered in the experiment, as described in the section entitled "additionnal STS measurements in superconducting regions" above. The various cases are described in detail in the caption of the figure.

\section{additionnal STS measurements in SDW regions}
\label{add_sdw_measurements}

\begin{figure}[tbh]\begin{center}\includegraphics[width=\columnwidth]{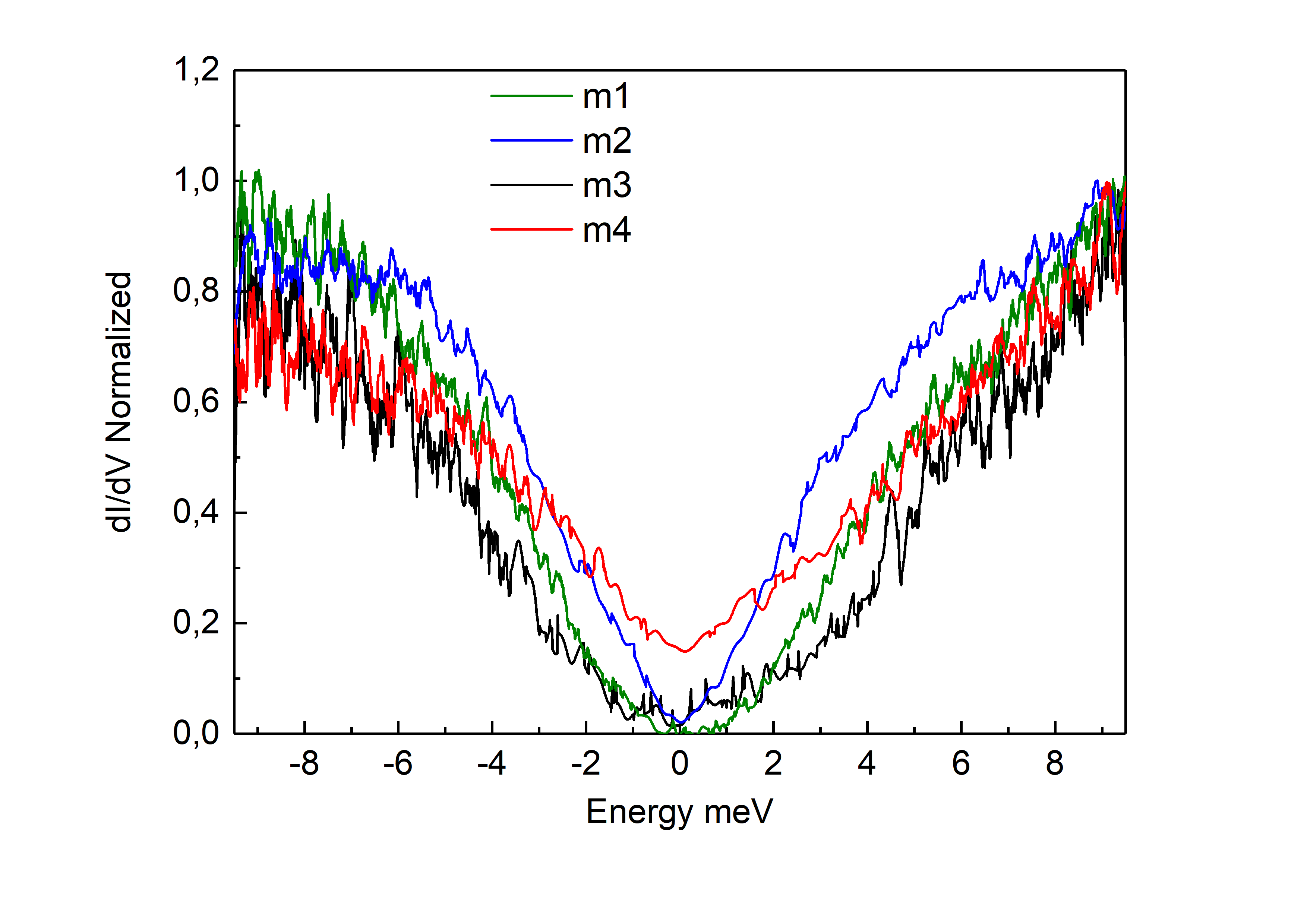}\caption{(color online) This figure summarizes $dI/dV$ excitation spectra associated with representative spin-density wave spectra measured at different locations and for different (TMTSF)$_2$ClO$_4$ samples from the same batch at $T=300$~mK. With respect to the terminology introduced in figure~\ref{fig_sts_sc_spectra}, all these spectra were acquired in "rough" regions. The typical set-point for spectroscopy was  $I=200$~pA to $I=300$~pA and $V=$-9~mV. \label{fig_sts_sdw_spectra}}\end{center}\end{figure}

The figure \ref{fig_sts_sdw_spectra} presents representative $dI/dV$ differential conductance spectra corresponding to spin-density wave excitation spectra. The various spectra were acquired at different locations and for different (TMTSF)$_2$ClO$_4$ samples from the same batch at $T=300$~mK.

\section{additionnal STS measurements showing the local variations of the high energy background}
\label{add_high_energy_measurements}

\begin{figure*}[tbh]\begin{center}\includegraphics[width=\textwidth]{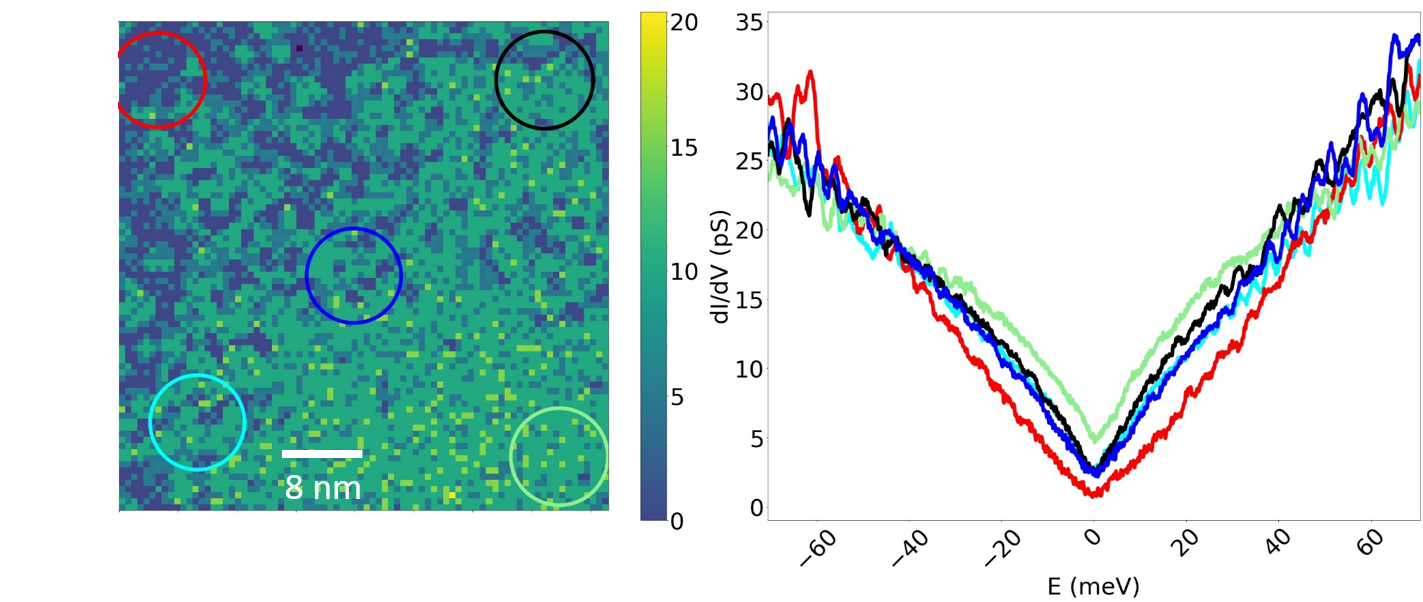}\caption{(color online) Left panel: $dI/dV(E)$ map measured at -20~meV over a 60$\times$60 nm$^2$ area. The colorbar is expressed in pS and is located on the right of the map. The STM topography (not shown here), acquired while performing this spectroscopy grid, was measured using $I=$~50~pA and $V_{bias}=-70$~mV. The set-point for each 75$\times$75 STS spectroscopy was $I=$~200~pA and $V_{bias}=-70$~mV, from which single $I(V)$ curves were acquired between [-70;+70]~mV a each point of the 75$\times$75 grid. Right panel: selected $dI/dV(E)$ spectra. Each colored spectrum represents an average inside a circle of comparable size comprising about 150 individual spectra and is indicated on the left $dI/dV(E)$ map using a corresponding color. The various observed high-energy backgrounds range from almost linear to convex.  \label{fig_high_energy_bckd}}\end{center}\end{figure*}

We also performed local measurements to estimate the local variations of the high-energy background encountered at the sample surface. To this end, full $I(V)$ maps were acquired at several areas on various samples between typically [-60;60]~mV. One example is shown in figure~\ref{fig_high_energy_bckd}. Local variations of the $dI/dV(E)$ background are measured across this area, exemplified by the selected spectra. The red spectrum, acquired in the top left part of the area presents characteristics very similar to the one presented in Fig.~\ref{fig3_STS_spec} of the main text, i.e. a close to linear energy dependence. The other spectra present an energy dependence closer to a power-law behavior with an exponent smaller than one. The microscopic origins of this feature is discussed in details in the main text. The local variations observed here at the nanometer scale are likely to be linked to structural disorder in the anions lattice. 

\section{Model of disordered DOS background}
\label{model_disord_DOS_bckgd}

The DOS in the  intermediate  energy range is governed by disorder coupled to electron-electron interaction. Following  Shinaoka and Imada \cite{Shinaoka09b,Shinaoka09}, it can be modeled by the Anderson-Hubbard model which consists of tight-binding electrons interacting through a local Hubbard  term $U$ between electrons of opposite spins and one-body  term of electrons that scatter on a spatially distributed random site potential $V$. From the numerical solution for  site-dependent Hartree-Fock equations, the energy profile of the DOS can be shown to scale according to the following expression in two dimensions:
\begin{equation}
    N_{\mathrm{AH}}(E) = c \exp \big[-\alpha  \ln^2(E  /t)\big]
\end{equation}
where $c$  and $\alpha$ are non universal constants. Here the energy difference $E$ is the energy difference from the Fermi energy $E_F$. For quasi-1D weakly dimerized metallic  chains at quarter-filling like the Bechgaard salts, the normalization hopping energy $t$  is connected to the  Fermi energy  by the relation $E_F= \sqrt{2}t$. The above expression is used to fit the STM data at intermediate energy.

\section{The density of states and the  quasi-1D electron gas model}
\label{model_Q1D_DOS}

The quasi-1D electron gas model for a linear array of $N_\perp$ weakly coupled  chains  consists of a non interacting hamiltonian with  the one-electron spectrum
\begin{equation}
\label{ep}
    E_p(\vec{k}) = v_F(pk-k_F) -2t_\perp\cos k_\perp -2t_\perp' \cos 2k_\perp,
\end{equation}
where $v_F$ is the Fermi velocity of right ($p=+$) and left ($p=-$) moving electrons along the chains ($\hbar =1$); $t_\perp$ and  $t_\perp'$ are  the transverse hopping between nearest and next-to nearest neighbor chains, respectively (interchain distance $d_\perp=1$). The hopping parameter $t_\perp'$, also called antinesting,  models deviations from perfect nesting conditions  at wavevector $\vec{Q}_0=(2k_F,\pi)$ for an open Fermi surface, namely 
\begin{equation}
\label{antin}
    E_+(\vec{k} +\vec{Q}_0) =-E_-(\vec{k}) -4t_\perp'\cos 2k_\perp.
\end{equation}
Regarding interactions  of the model these are commonly defined by three main scattering amplitudes on the Fermi surface, that is, the backward and forward normal scattering  processes $g_{1,2}(k_{\perp1}',k_{\perp2}';k_{\perp1},k_{\perp2}) $ for large ($\sim 2k_F$) and small ($\sim 0$) momentum transfer along the chains and  $g_{3}(k_{\perp1}',k_{\perp2}';k_{\perp1},k_{\perp2}) $ for umklapp scattering that transfers two electrons from  one side  ($p$) of the Fermi surface to the other ($-p$). By lowering energy,  the couplings acquire a momentum dependence along the Fermi surface parameterized  by  ingoing ($k_{\perp1,2}$) and outgoing ($k_{\perp1,2}'$) transverse wave vectors.

The possible instabilities of the normal state against the formation of either an ordered SDW  or a SC state at low energy can be obtained by the renormalization group method \cite{Sedeki12,Bourbon09,Nickel06}.  At the one-loop level, the  method proceeds to the successive integration of electron degrees of freedom as a function of energy distance $E$ from the Fermi surface. It allows to sum   electron-hole (e-h) closed  and vertex corrections diagrams with interfering   electron-electron (e-e) ladder diagrams. This yields the flow equations of the normalized scattering amplitudes \cite{Sedeki12}, which can be written in the  schematic form   
  \begin{align}
\label{gn}
 \!\!  \!\! E\,\partial_E     {g}_{i=1,2}(\{k_\perp\}) \!   = &\sum_{{\cal P}_{nn'}}\!\Big\{\epsilon^{n,n'}_{C,i}\! \!\left\langle {g}_n  \cdot {g}_{n'} \cdot E \partial_E {\cal L}_C\right\rangle_{k_\perp}  \cr      
   &  +   \  \epsilon^{n,n'}_{P,i}\!\left\langle    {g}_n  \cdot {g}_{n'} \cdot  E \partial_E{\cal L}_P\right\rangle_{k_\perp} \! \Big\}, \cr\cr
 \!\!  \!\!  E\,\partial_E  {g}_3(\{k_\perp\})      =  &  \sum_{{\cal P}_{3n}} \epsilon^{3,n}_{P,3} \,\left\langle {g}_3 \cdot {g}_{n}\cdot  E \partial_E{\cal L}_P\right\rangle_{k_\perp}, \cr
\end{align}
where the  $k_\perp$-dependence of  convolution products, constrained by momentum conservation,  has been  masked  for simplicity; here $\langle \ldots \rangle_{k_\perp}$ stands for an average over $k_\perp$. The e-h (P) and e-e (C) loops at finite $T$ are given by
\begin{eqnarray}
&&\!\!\! E \partial_E{\cal L}_{P,C}(k_\perp, q_{P,C})=  -\sum_{\nu=\pm} \theta(|E + \nu {\cal A}_{P,C}|- E )\cr
&& \times {1\over 2}\left( \tanh{E  + \nu {\cal A}_{P,C}\over 2k_BT} + \tanh{ E\over 2k_BT}\right)\cr
&& \times {E\over 4E +  \nu {\cal A}_{C,P}},
\end{eqnarray}
 which  are  evaluated at $q_P = k_{\perp1}-k_{\perp1}'$ and $q_C=k_{\perp 1}+k_{\perp 2}$, respectively; ${\cal A}_{P,C}(k_\perp,q_{P,C})= -E_+(k_F,k_\perp) \mp E_-(-k_F,k_\perp + q_{P,C})$ and   $\theta(x)$ is the step function.  For a given $g_i$, the related sum  in (\ref{gn}) collect  all  possibilities ${\cal P}_{nn'}$  of diagrams with normal scattering $g_{n=1,2}$ $(g_{n'=1,2})$ and those ${\cal P}_{3,n}$  involving umklapp scattering.     The coefficients $\epsilon_{C,P,i}^{nn'}$ fix  the   sign and spin multiplicity  of each loop contribution; it differs for closed loops    ($\epsilon_{P,i}^{nn'}=-2)$, vertex corrections ($\epsilon_{P,i}^{nn'}=1)$ and ladder  ($\epsilon_{C,i}^{nn'}=-\epsilon_{P,i}^{nn'}=-1$) diagrams.

Substituting  in these flow equations initial parameters that reasonably describe the Bechgaard salts, the instabilities of the metallic  state can then be followed  as function of pressure simulated by $t_\perp'$.     For band parameters, one has typically $v_Fk_F = E_F\simeq k_B 3000$~K for the high  (Fermi)  energy cutoff and $t_\perp/k_B = 200$~K  for the smaller transverse hopping amplitude. As for the couplings different experiments can fix their  initial normalized  values  at $E=E_F$, with   $ g_1 \simeq 0.3 $ and $g_2 \simeq 0.6$  and  $g_3\simeq .03$ for the small, but essential, half-filling umklapp amplitude, a consequence of the weak dimerization of molecular chains  in the Bechgaard salts. 

 \begin{figure*}[tbh]\begin{center}\includegraphics[width=\columnwidth]{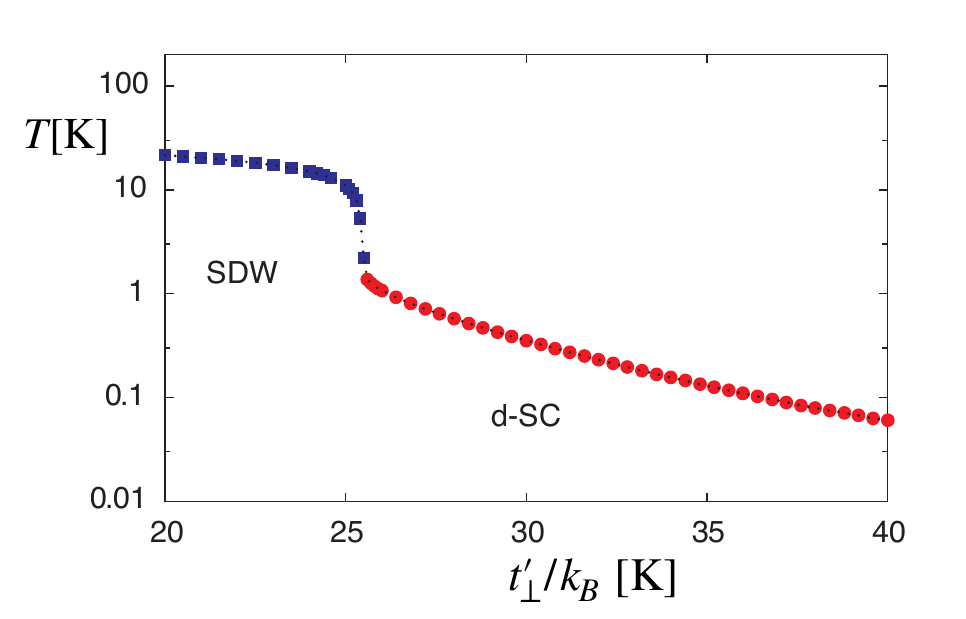}\caption{Calculated RG phase diagram for the quasi-one dimensional electron gas model of the Bechgaard salts.
 \label{PhasesD}}\end{center}\end{figure*}
 
Following the integration of (\ref{gn}) down to $E\to0$ at the Fermi surface, two possible instabilities can occur  as a function of $t_\perp'$ at a given critical temperature scale. The first  is against the formation of a SDW state at the  ordering temperature $T_{\mathrm{SDW}}$ which is signaled by a singularity in the  combination of couplings $(g_2+g_3)_{\{k_\perp,\vec{Q}_0\}}$, for $k_\perp$ along the Fermi surface and at the momentum transfer $\vec{Q}_0$ of the SDW modulation. Increasing $t_\perp'$,  $T_{\mathrm{SDW}}$ undergoes a monotonic decrease until a critical $t_\perp'^*$ is approached from below     and where a sharp drop of  $T_{\mathrm{SDW}}$  points towards a quantum critical point (QCP). However, the contact with the QCP  at zero temperature is prevented  by another singularity at temperature $T_c$, occuring this time in the combination $(g_1+g_2)_{\{k_\perp,\vec{Q}_C\}}$ at zero pair momentum $\vec{Q}_C=0$ for ingoing and outgoing electrons.  The combination of couplings is cosine modulated in  $k_\perp$ space. This signals an instability towards the formation of   a  d-wave superconducting  gap $\Delta(k_\perp) = \Delta \cos k_\perp$ along the Fermi surface below  $T_c$.   Moving away from the QCP as $t_\perp'$ is increased,  $T_c$ declines progressively. This  characteristic sequence of instabilities as a function of the pressure parameter $t_\perp'$ is summarized in  the  phase diagram of Figure~\ref{PhasesD}.

\begin{figure*}[tbh]\begin{center}\includegraphics[width=\columnwidth]{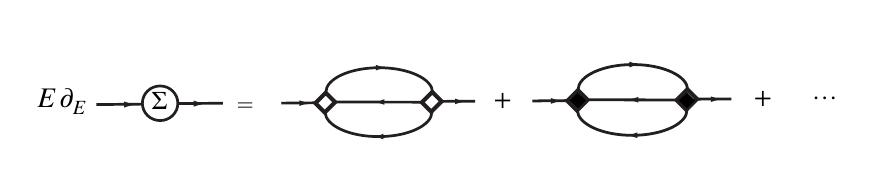}\caption{\label{Self}  One-particle self-energy RG flow equation at the two-loop $\mathcal{O}(g^2)$ level. The open (full) diamond stands for normal (umklapp) electron-electron scattering vertex. Adapted from Ref.~\cite{Sedeki12}.  }\end{center}\end{figure*}

The impact of the above one-loop scattering amplitudes on quasi-particle spectral properties of the model can be obtained from the calculation of one-particle self-energy. The corresponding  flow equation  derived at the two-loop level in Ref.~\cite{Sedeki12} is shown in the diagrammatic form in Fig.~\ref{Self}.

From that work, the  quasi-particle weight $z(k_\perp,E)$ located at $\vec{k}_F(k_\perp)$ for $p$  electrons on the Fermi surface and  energy distance $E$ can be written in the  schematic  form  
\begin{equation}
  \begin{split}
  \label{zeq}
  &E\,\partial_E \ln z({k}_{\perp},E) =   \int \frac{d^2k'_\perp}{(2\pi)^2}\cr
& \times  \Big\{ \big( g_2\cdot g_1 -g_2\cdot g_2-g_1\cdot g_1 \big)\,\cdot I_{n,E}\cr
  &- g_3\cdot g_3
\cdot I_{3,E}\Big\},
\end{split}
\end{equation}
in which the transverse momentum dependence and diagram permutations have been masked for simplicity. Here the transverse momentum and energy dependent expressions of the two-loop diagrams of normal and umklapp  vertices are of the form  
\begin{equation*}
    I_{n(3),E} \sim \frac{1}{2} \frac{T^2}{L^2} (\pi v_F)^2 \partial_E\sum_{\{\omega\}}\int_E^{E_F} dE'G_{-p,E}^0\cdot G_{-p,E'}^0\cdot G_{p,E'}^0
\end{equation*}
where $G_{p,E}^0= [i\omega_n -E_p(\vec{k})]^{-1}$ is the bare electron propagator for the $p$ branch.

Single particle tunneling along the $c^*$ direction probed by STM  experiments couples to local density of states derived from the quasi-particle weight of each chain of the conducting $ab$ plane, namely $N_n(E) = N_0 \langle z(k_\perp,E)\rangle_{k_\perp}$, where $N_0=2(\pi v_F)^{-1}$ is the total bare density of states for both spin orientations.

\begin{figure*}[tbh]\begin{center}\includegraphics[width=\columnwidth]{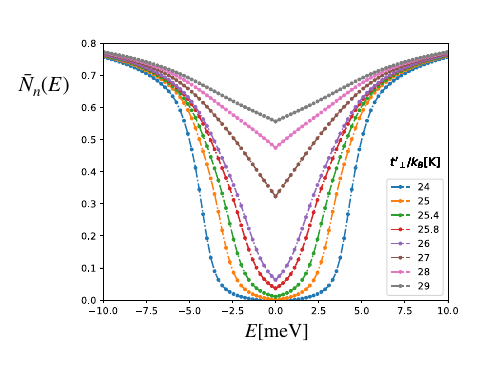}\caption{ \label{NvsE} Energy dependence of the normal state single particle density of states for the quasi-1D electron gas model as calculated by the RG method  for different antinesting $t_\perp'$ values of the phase diagram of Fig.~\ref{PhasesD}. The temperature   is set  above but close  the critical point $T\gtrsim  T_{\mathrm{SDW},c}$. }\end{center}\end{figure*}

\begin{figure*}[tbh]\begin{center}\includegraphics[width=\columnwidth]{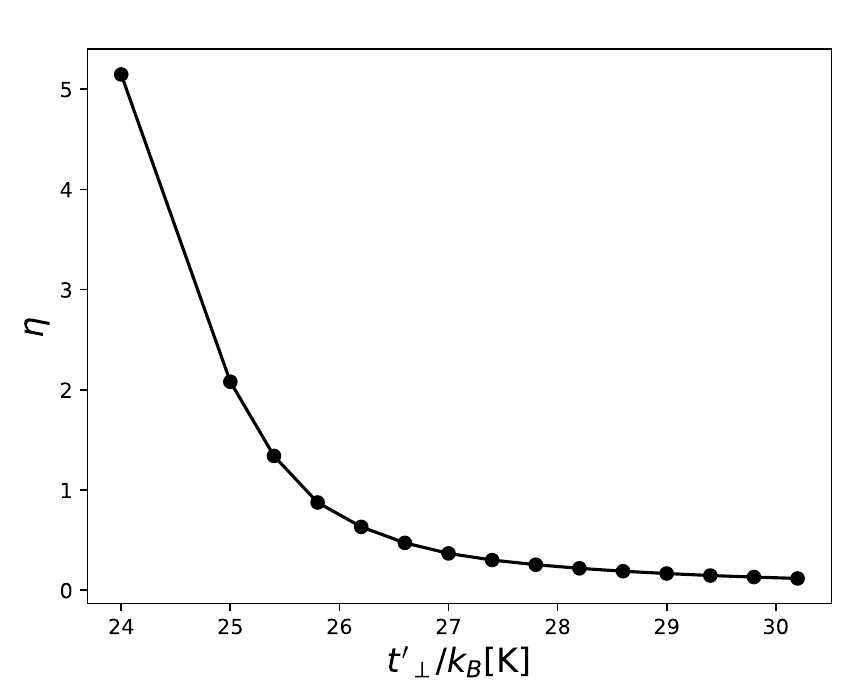}\caption{\label{eta}Variation of the power law exponent for the calculated single-particle density of states in the antinesting low energy range ($E<E_0^*$) and for different $t_\perp'$. }\end{center}\end{figure*}

Inserting the results for  the scattering amplitudes (\ref{gn}) in the  flow equation (\ref{zeq}) for self-energy, one finds in Fig.~\ref{NvsE} the energy profile of  density of states of the metallic phase; it is    computed at  different antinesting $t_\perp'$  and    for a temperature $T$ above but close the critical  $T_{c,\mathrm{SDW} }$. The results display a  characteristic dip at the Fermi level with a  depth and energy dependence that vary with $t_\perp'$. According to (\ref{antin}),  the latter indeed introduces an energy scale at $E_0^*\sim 2t_\perp'$ ($\sim 5$meV for the model parameters used in Fig.~\ref{NvsE})  below which  the electron-hole pairing responsible for  SDW correlations becomes $k_\perp$-dependent and  then   weakened. It follows from   (\ref{gn}) an interplay between SDW and d-wave SC correlations at $E< E_0^*$, which is responsible  for a dip in $N_n(E)$ that is well described by a power law  in energy:
\begin{equation}
    \bar{N}_n(E) \simeq C \big(\mathrm{Max}\{\pi k_BT,|E|\}\big)^\eta, \ \ \ \ \ (E<E_0^*\sim 2t_\perp').
\end{equation}
which is cutoff by the temperature at sufficiently low energy. The  exponent $\eta$ and constant $C$ are   $t_\perp'$  dependent as shown in Fig.~\ref{eta}. At  the left of the QCP, on the SDW side  of the phase diagram, $\eta$  grows above unity which is  compatible with a concave U-shape of $N_n(E)$  with energy. Close to the QCP on the SC side,   $N_n(E)$ displays a   linear V-shape  energy dependence with $\eta\simeq 1$ which  transforms into a convex variation as $\eta$  progressively decreases below unity when $t_\perp'$ is made larger.

\section{Calculated tunneling conductance in the ordered state}
\label{model_tunnel_cond_ord_state}

The above perturbative renormalization group results for the energy dependent density of states are restricted to the normal  metallic  state. The impact of the entry  into the ordered phase at low energy  will be accounted for by completing the expression of the density of states by a  BCS form for $|E|< \Delta_\mu$ and   in the low temperature limit \cite{Dynes78,Suzumura89}. One shall consider the expression
\begin{equation}
\label{Ns}
\bar{N}(E)= \bar{N}_n(\Delta_\mu,|E|)\Big\langle{\rm Re}\Bigg\{ {|E+ i\Gamma| \over \sqrt{ (E+ i\Gamma)^2 - \Delta_\mu^2(k_\perp)}}\Bigg\} \Big\rangle_{k_\perp},
\end{equation}
where ${\bar{N}_n(\Delta_\mu,|E|) =C (\text{Max}\{\Delta_\mu,|E|\})^\eta }$ is the normal part. Here the gap $\Delta_\mu(k_\perp)$ is given by    $\Delta_{\rm SC}(k_\perp)=\Delta\cos k_\perp$ for a quasi-1d d-wave superconductor and by the constant  $\Delta_{\rm SDW}(k_\perp)=\Delta$ in the insulating SDW case; $\Gamma$ stands for the impurity scattering rate that  broadens  the coherence peaks for both orderings.

\begin{figure*}[tbh]\begin{center}\includegraphics[width=\columnwidth]{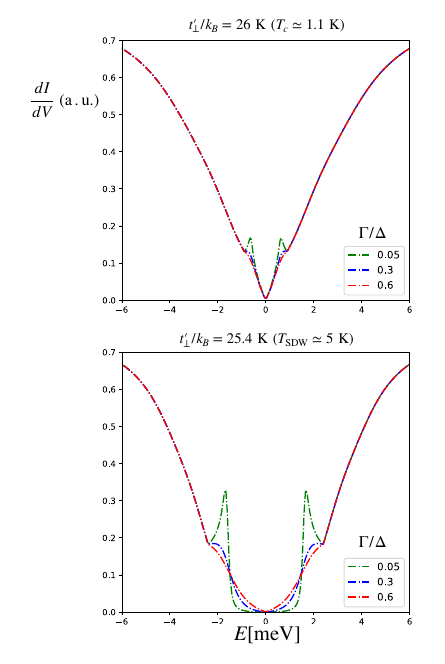}\caption{\label{conducRG}Low energy dependence of the calculated  differential conductance [Eq. (\ref{dIdVCal})] at $T\ll T_\mu$ in (top) the ordered SC ($t_\perp'/k_B=26$~K, $T_c\simeq 1.1$~K) and (bottom) SDW ($t_\perp'/k_B=25.4$~K,  $T_{\mathrm{SDW}}\simeq 5$~K) phases at different normalized scattering rate parameters  $\Gamma/\Delta$.} \end{center}\end{figure*}

The above density of states  takes place in the expression of the differential conductance  
\begin{equation}
\label{dIdVCal}
{dI\over dV}(E) =   G_{nn}\int_{-\infty}^{+\infty} \bar{N}(\omega)\big[ -{\partial f(\omega+ eV)\over \partial (eV)}\big] d\omega,
\end{equation} 
where $E=eV$, $f(\omega)$ is the Fermi distribution  and $G_{nn}$ is adjustable  constant  for the amplitude of  normal conductance. The low energy dependence of $dI/dV$ is illustrated  in Figure~\ref{conducRG}  for two typical values of pressure $t_\perp'$ in the SC and SDW domains of the phase diagram. In the SC case, one sees the suppression of the coherence peaks in the density of states by the impact of the scattering rate. The comparison with the experimentally measured spectra supports high scattering rates occurring at the sample surface, most probably linked to the cleavage process and disorder in the $ClO_4$ ions sublattice. In the SDW case where the reduction of the normal part $\bar{N}_n$ is relatively more pronounced, this effect makes barely visible the impact of the ordered phase on the density of states. Nevertheless, a close look at figures~\ref{STMRG}a) and \ref{STMRGSDW} of the main text show small kink features located around 1~meV and 2.5~meV, respectively, that are consistent with the theoretically expected superconducting and SDW peaks broadened by disorder shown in Fig.~\ref{conducRG} of the present supplementary material.\newline\newline

*christophe.brun@sorbonne-universite.fr, Claude.Bourbonnais@USherbrooke.ca

\end{document}